\documentclass[graybox, envcountchap]{svmult}

\usepackage{mathptmx}        
\usepackage{amsmath}
\usepackage{amssymb}
\usepackage{color}
\usepackage{helvet}          
\usepackage{courier}         
\usepackage{dirtree}
                             
\usepackage{makeidx}        
\usepackage{graphicx}        
\usepackage{subfig}

\usepackage{multicol}        
\usepackage[bottom]{footmisc}

\usepackage{hyperref}        
\hypersetup{colorlinks=true,urlcolor=blue}

\usepackage[misc]{ifsym}

\makeindex             

\begin{document}
\title{On the interacting dark energy scenarios $-$ the case for Hubble constant tension}
\author{Supriya Pan and Weiqiang Yang}
\institute{Supriya Pan\\Department of Mathematics, Presidency University, 86/1 College Street, Kolkata-700073, India\\ \email{supriya.maths@presiuniv.ac.in}\\\\
Weiqiang Yang\\Department of Physics, Liaoning Normal University, Dalian, 116029, P. R. China\\ \email{d11102004@163.com}}
%
\maketitle
\abstract{The Hubble constant $H_0$ is one of the important cosmological parameters measuring the expansion rate of our universe at present moment. Over the last couple of years, $H_0$ has created an enormous amount of debates interests in the astrophysical and cosmological communities for its different estimations at many standard deviations by different observational surveys. The recent estimation of $H_0$ from the cosmic microwave background observations by Planck within the $\Lambda$-Cold Dark Matter ($\Lambda$CDM) paradigm is  in tension with $\gtrsim 5 \sigma$ confidence with SH0ES (Supernovae and $H_0$ for the Equation of State of dark energy) collaboration. As a result, revision of the standard $\Lambda$CDM model has been suggested in various ways in order to examine whether such scenarios can solve this $H_0$ tension. Among the list of the proposed cosmological scenarios, in this chapter we focus on a generalized cosmological theory in 
which the dark components of the universe, namely, Dark Matter (DM) and Dark Energy (DE) are allowed to interact with each other in a non-gravitational way, widely known as the Interacting DE or Coupled DE scenarios. These interacting scenarios have received magnificent attention in the scientific community for their appealing consequences. Specifically, in the context of $H_0$ tension, it has been observed that the interacting DE scenarios can lead to higher values of the Hubble constant ($H_0$) value, and consequently, the tension on $H_0$ can be either alleviated or solved. In this chapter we review various interacting DE scenarios and their roles in alleviating the $H_0$ tension.  }

\newpage

\section{Introduction}
\label{sec:1}

The dark sector of our universe is highly mysterious and it has become a very appealing topic for modern cosmology and  cosmologists. According to a large amount of observational evidences, the dark sector of the universe is comprised of two heavy fluids, namely, 
dark matter (DM) and dark energy (DE) together which occupy nearly 96\% of the total energy budget of our universe. The DM sector is responsible for the observed structure formation of our universe while the role of DE is to drive the accelerating expansion of the  universe \cite{SupernovaSearchTeam:1998fmf,SupernovaCosmologyProject:1998vns}. While the origin, nature and evolution of this dark picture is not fully understood at this moment, but on the report of a large number of astronomical observations, 
$\Lambda$-Cold Dark Matter model, a very simple cosmological model constructed in the context of Einstein's General Relativity (GR) where $\Lambda >0$ acts as the source of DE and DM being pressure-less, has been quite successful in describing the current observational evidences. In this simplest cosmological picture, DE and DM are conserved independently, that means, evolution of DM does not affect the evolution of DE, and vice-versa. 
Despite marvellous successes, $\Lambda$CDM cosmological model has some limitations. The cosmological constant problem is one of the biggest issues that still needs an answer \cite{Weinberg:1988cp}.  On the other hand,  the cosmic coincidence problem (the so-called ``why now'' problem) \cite{Zlatev:1998tr} is another cosmological issue associated with this cosmological model. Apart from the above issues, it was reported that the estimations of some key cosmological parameters using the cosmic microwave background (CMB) observations by Planck (within the $\Lambda$CDM paradigm) are in tension at many standard deviations with the local measurements.  In particular, the measurements of the Hubble constant by the cosmic microwave background observations by Planck within the $\Lambda$CDM paradigm ($H_0 = 67.4 \pm 0.5$ km/s/Mpc at 68\% CL)~\cite{Planck:2018vyg} and SH0ES (Supernovae and $H_0$ for the Equation of State of dark energy) collaboration ($H_0 = 73.0 \pm 1.0$ km/s/Mpc at 68\% CL, labeled as R21)~\cite{Riess:2021jrx} are in tension with $\gtrsim 5 \sigma$ confidence. Additionally, some further estimations by the SH0ES team indicate that this tension can increase up to $5.3\sigma$ \cite{Riess:2022mme}.  
Moreover, the $S_8$ parameter, defined as a combination of the amplitude of the matter power spectrum $\sigma_8$ with the matter density at present $\Omega_m$ ($S_8=\sigma_8\sqrt{\Omega_m/0.3}$) exhibits a tension 
at the level of $2-3\sigma$ between Planck (within $\Lambda$CDM paradigm)
and low redshift probes, e.g. weak gravitational lensing and galaxy clustering (e.g.~\cite{Asgari:2019fkq,KiDS:2020suj,Joudaki:2019pmv, DES:2021wwk,DES:2021bvc, DES:2021vln, KiDS:2021opn,Heymans:2020gsg,DES:2020ahh,Philcox:2021kcw}). Thus, even if $\Lambda$CDM is extremely successful with a series of observational data, however, based on the aforementioned limitations, it is argued that perhaps $\Lambda$CDM is an approximation of a more realistic scenario that is yet to be discovered. Following this, several modifications of the $\Lambda$CDM cosmological scenario were proposed by many authors, see the recent reviews \cite{DiValentino:2021izs,Perivolaropoulos:2021jda,Abdalla:2022yfr,Kamionkowski:2022pkx}.

In this chapter we focus on a very generalized cosmological scenario where  DE and DM interact with each other in a non-gravitationally way. That means an exchange of energy and/or momentum between DE and DM is allowed.
This class of cosmological models is known as Interacting DE (or coupled DE or Interacting DE-DM). According to the historical records, the idea of interaction between gravity and a scalar field with exponential potential was proposed by Wetterich to explain the tiny value of the cosmological constant \cite{Wetterich:1994bg}. However, in Wetterich's proposal \cite{Wetterich:1994bg}, explicit DE-DM interaction was not introduced. Soon after the discovery of the late-time accelerating expansion of our universe, when the dynamical DE in the name of quintessence was proposed and the cosmic coincidence problem emerged in the cosmological domain \cite{Zlatev:1998tr}, Amendola wrote an article on the ``Coupled quintessence'' \cite{Amendola:1999er} in which he formally introduced an interaction between the matter sector and a scalar field with exponential potential acting as a source of DE where the author pointed out that such an interaction  can solve the long debating ``cosmic coincidence'' problem. This coincidence problem was further investigated by other authors and it was found that interacting DE models can excellently solve the cosmic coincidence problem ~\cite{Chimento:2003iea,Cai:2004dk,Pavon:2005yx,delCampo:2006vv,Berger:2006db,delCampo:2008sr,delCampo:2008jx}. 
This was the first achievement of the theory of interacting DE and this influenced various investigators to work with the interacting DE models. As a result, over the last several years a number of interacting DE models have been investigated with some interesting outcomes ~\cite{Barrow:2006hia,Amendola:2006dg,Koivisto:2006ai,He:2008tn,Gavela:2009cy,Majerotto:2009np,Valiviita:2009nu,Clemson:2011an,Salvatelli:2014zta,Yang:2014vza,Yang:2014gza,Wang:2014xca,Faraoni:2014vra,Pan:2012ki,Yang:2014hea,Li:2015vla,Nunes:2016dlj,Yang:2016evp,Pan:2016ngu,Mukherjee:2016shl,Sharov:2017iue,Guo:2017hea,Cai:2017yww,Yang:2017yme,Yang:2017ccc,Yang:2017zjs,Pan:2017ent,Yang:2018pej,Yang:2018ubt,Paliathanasis:2019hbi,Pan:2019jqh,Yang:2019bpr,Yang:2019vni,Yang:2020zuk,Pan:2020bur} (also see the review articles on the interacting DE scenarios \cite{Bolotin:2013jpa,Wang:2016lxa}). 
In particular, it was observed that an interaction between DE-DM can lead to a phantom DE scenario 
\cite{Wang:2005jx,Das:2005yj,Sadjadi:2006qb,Pan:2014afa} without invoking any scalar field theory with negative kinetic term which may lead to instabilities in classical and quantum levels.  
In the context of the cosmological tensions, interacting DE plays a very promising role. There are many evidences showing that an interaction between DE and DM can increase the $H_0$ value and thus such scenarios can alleviate the $H_0$ tension between Planck (within the $\Lambda$CDM paradigm) and SH0ES collaborations \cite{Kumar:2016zpg,Kumar:2017dnp,DiValentino:2017iww,Yang:2018euj,Yang:2018uae,Kumar:2019wfs,Pan:2019gop,DiValentino:2019ffd,DiValentino:2019jae,Zhai:2023yny}. It has been also noticed that the interacting DE has the potentiality to alleviate the  $S_8$ tension \cite{Kumar:2017dnp,Kumar:2019wfs,Pourtsidou:2016ico,An:2017crg}. Based on the above observations, it is clear that the cosmological scenarios allowing a non-gravitational interaction between DE and DM are highly appealing and they should get more attention in the forthcoming years.

This chapter is organized in the following way. In section~\ref{sec2-basic-equations} we give an overview of the interacting DE-DM models and  present the evolution equations of a general interaction model at the level of background and perturbations.  In section \ref{sec-models} we discuss various models of interaction that have been investigated in the literature to alleviate the Hubble tension. In particular, in section \ref{sec-IDE-const-w} we discuss the interacting models in which the DE equation of state is constant; in section \ref{sec-IDE-variable-w} we present those results where the DE equation of state is dynamical; and in section \ref{sec-variable-coupling} we discuss a special kind of interaction models in which the coupling parameter of the interaction scenarios is assumed to be time dependent. 
In section \ref{sec-summary}
we conclude the chapter with a brief summary of the entire work.

\section{Interacting DE}
\label{sec2-basic-equations}

In the large scales, our universe is almost homogeneous and isotropic. This homogeneous and isotropic  configuration of the universe is  well described by the Friedmann-Lema\^{i}tre-Robertson-Walker (FLRW)  line element 
\begin{eqnarray}
	ds^2=-{dt}^2+ a^2(t) \left[\frac{dr^2}{1-kr^2} + r^2 (d\theta^2 + \sin^2 \theta d \phi^2) \right]
\end{eqnarray}
where ($t, r, \theta, \phi$) are the co-moving coordinates;  $a(t)$ is the expansion scale factor of the universe; $k$ denotes the spatial geometry of the universe which can take three distinct values, namely, $0, +1, -1$ corresponding to three distinct spatial geometries of the universe $-$ flat ($k =0$), closed ($k = +1$) and open ($k =-1$). 
We further assume that the gravitational sector of the universe is well described by GR and the matter distribution is minimally coupled to gravity. In the matter distribution, one can consider a variety of components, such as radiation, baryons, neutrinos, dark matter, dark energy. 
Now, within this framework, the Einstein's gravitational equations $G_{\mu \nu} = 8 \pi G\; T_{\mu \nu}$ can be recast as 
\begin{eqnarray}
H^2 + \frac{k}{a (t)^2} = \frac{8 \pi G}{3} \sum_{i} \rho_i, \label{efe-1}\\
2 \dot{H} + 3 H^2 + \frac{k}{a (t)^2} = - 8 \pi G \sum_{i} p_i,\label{efe-2}
\end{eqnarray}
where an overhead dot represents the derivative with respect to the cosmic time; $H \equiv \dot{a} (t)/a(t)$ is the Hubble rate of the FLRW universe; $p_i$ and $\rho_i$ are respectively the pressure and energy density of the $i$-th fluid. Hence $\sum_i \rho_i$ and $\sum_i p_i$ denote  the total energy density and total pressure of the matter distribution, respectively. 
Using the Bianchi's identity $\nabla^{\mu} T_{\mu \nu} = 0$, or, alternatively, using eqns. (\ref{efe-1}) and (\ref{efe-2}), one can derive the conservation equation of the total fluid as follows 
\begin{eqnarray}
\sum_i \dot{\rho}_i + 3 H \sum_i (\rho_i + p_i) = 0. \label{cons}
\end{eqnarray}
We assume that DM and DE are coupled to each other, but the remaining fluids are independently conserved. That means, in the context of the FLRW universe, the conservation equations for DM and DE become, 
\begin{eqnarray}
\dot{\rho}_{\rm dm} + 3 H (1+w_{\rm dm}) \rho_{\rm dm} = -Q (t),\label{cons-CDM}\\
\dot{\rho}_{\rm de} + 3 H (1+w_{\rm de}) \rho_{\rm de} = +Q (t),\label{cons-DE}
\end{eqnarray}
where $\rho_{\rm dm}$ and $\rho_{\rm de}$ are respectively the energy density of DM and DE; $w_{\rm dm} \geq 0$, $w_{\rm de} < -1/3$,  are respectively the barotropic equation of state parameters of DM and DE. The quantity $Q$ denotes the rate of energy or/and momentum exchange between these dark fluids and it is the heart of the theory of interacting dark energy. For $Q (t) > 0$, the energy flows from DM to the DE while for $Q (t) <0$, energy flow takes place from DE to DM.
In the limiting case $Q \rightarrow 0$, the non-interacting cosmological scenario as a special case is recovered. Now, given a specific interaction function, one can in principle determine the dynamics of the universe at the background level using the conservation equations (\ref{cons-CDM}), (\ref{cons-DE}) together with the Hubble equation (\ref{efe-1}). The interaction function also affects the evolution of the universe at the perturbation level. 
However, there is no such guiding principle to select the interaction function. Although several attempts have been made to derive the interaction function from some action~principle~\cite{Gleyzes:2015pma,Boehmer:2015kta,Boehmer:2015sha,vandeBruck:2015ida,Xiao:2018jyl,Amico:2016qft,Marsh:2017prd,Kase:2019hor,Alexander:2019wne,Pan:2020zza}, but no decisive conclusion has been made yet. Thus, in most of the cases, the choice of the interaction function is made from the phenomenological ground and hence there is no unique choice of the interaction function. The existing interaction functions are usually divided into two classes:  the interaction functions including explicitly the Hubble rate $H$ of the FLRW universe, e.g. $Q (t) = H f(\rho_{\rm dm}, \rho_{\rm de})$, where $f$ is an analytic function of the energy densities of the dark components. A general form of the interaction function in this category that recovers many interactions is the following 
$Q = 3 H \xi \rho_{dm}^{\alpha} \rho_{de}^{\beta} (\rho_{dm}+ \rho_{de})^{\gamma}$,
where $\alpha + \beta + \gamma = 1$ and $\xi$ is a dimensionless coupling parameter. Note that for different values of $(\alpha, \beta, \gamma)$, one can produce a cluster of interaction functions with only one coupling parameter. The well known interaction function $Q = 3 H \xi \rho_{de}$ is obtained for $\alpha = 0 = \gamma$. We stress that the above choice is a generalized interaction function in a specific direction and this does not include other kind of interaction models, e.g. the interaction models with two  coupling parameters, see for instance \cite{Pan:2017ent,Pan:2016ngu}, or the interaction functions involving the exponential or trigonometric functions \cite{Pan:2020bur}. 
Secondly,  one can consider another class of interaction functions where the Hubble rate does not appear explicitly, i.e. of the form $Q (t) = f(\rho_{\rm dm}, \rho_{\rm de})$.  A generalized interaction function can be of the following: 
$Q = \Gamma  \rho_{dm}^{\alpha} \rho_{de}^{\beta} (\rho_{dm}+ \rho_{de})^{\gamma}$,
where $\alpha + \beta + \gamma = 1$ but here $\Gamma$ has the dimension equal to the dimension of the Hubble parameter.\footnote{In the statistical analysis, we thus consider the dimensionless parameter $\Gamma/H_0$.}
It has been argued in many places, see for instance, Ref. \cite{Valiviita:2008iv}
that as the interaction between DE and DM is a local phenomenon,  therefore, it should depend only on the local quantities and the  appearance of the Hubble parameter $H$, the global expansion rate, in the interaction functions is mainly designed for the mathematical simplicity. While as argued in Ref. \cite{Pan:2020mst}, explicit appearance of $H$ in the interaction functions can be motivated from some well established cosmological theories. Thus, the mathematical simplicity does not strongly hold in such cases. On the other hand, it has been observed in Ref. \cite{Yang:2021oxc} that a particular choice of the interaction function may lead to negative energy density of DM or (and) DE component either in the past or in the future. This is a striking issue since the consequences of the negative energy density of the dark components are not clearly understood at this moment.

Notice that depending on the nature of  the equation of state parameters, namely, 
$w_{\rm dm}$, $w_{\rm de}$, various interacting scenarios can be considered. 
In most of the cases, $w_{\rm dm}$ is set to zero due to the abundances of the cold DM in the universe sector\footnote{However, as the nature of DE and DM is still yet to be discovered,  and since the sensitivity of the observational data is increasing over time,  therefore, there is certainly no guiding rule to consider $w_{\rm dm} = 0$. In fact, one may consider  $w_{\rm dm}$ to be a free parameter in the underlying cosmological scenarios and let the observational data to decide its fate, see for instance, Refs. \cite{Muller:2004yb,Kumar:2012gr,Armendariz-Picon:2013jej,Kopp:2018zxp,Ilic:2020onu,Naidoo:2022rda,Pan:2022qrr}. }, and the nature of $w_{\rm de}$ remains unspecified, that means it can be either  constant or dynamical. 
Notice that the conservation equations (\ref{cons-CDM}) and (\ref{cons-DE}) can alternatively be expressed as 
\begin{eqnarray}
\dot{\rho}_{dm} + 3 H (1+ w^{\rm eff}_{dm}) \rho_{dm} = 0, \label{cons-CDM-1}\\
\dot{\rho}_{de} + 3 H (1+ w^{\rm eff}_{de}) \rho_{de} = 0,\label{cons-DE-1}
\end{eqnarray}
where $w^{\rm eff}_{de}$ and $w^{\rm eff}_{dm}$ are termed as the effective equation-of-state parameters of DM and DE, and they are given by
\begin{eqnarray}
&& w^{\rm eff}_{dm} =  w_{\rm dm} + \frac{Q (t)}{3H \rho_{\rm dm}}. \label{eff-eos-dm}\\
&& w^{\rm eff}_{de} = w_{\rm de} - \frac{Q (t)}{3H \rho_{\rm de}}. \label{eff-eos-de}
\end{eqnarray}
Thus, we see that even though we started with an interacting scenario between DE and DM quantified through the interaction function $Q (t)$, but this interacting scenario can always be expressed as a non-interacting scenario described by the conservation equations in 
eqns. (\ref{cons-CDM-1}) and (\ref{cons-DE-1}) in terms of two effective equation of state parameters of DM and DE displayed in eqns. (\ref{eff-eos-dm}) and (\ref{eff-eos-de}). Now, if the DM sector is cold, i.e. $w_{\rm dm} = 0$, then  $w^{\rm eff}_{dm} = \frac{Q (t)}{3H \rho_{\rm dm}}$. 
We remark that the presence of $Q (t)$ in eqns. (\ref{eff-eos-dm}) and (\ref{eff-eos-de}) can significantly affect their nature. For example, in an expanding universe ($H > 0$), an interacting quintessence model in which $w_{\rm de} > -1$ can offer an effective phantom-like behaviour ($w_{\rm de} < -1$) provided $Q (t) >0$.  This is indeed an interesting observation since  a phantom equation-of-state in DE can increase the expansion rate of the universe faster and as a result, the Hubble constant can take higher values~\cite{DiValentino:2016hlg}. 
In this context, an interaction which can offer a phantom DE equation of state through (\ref{eff-eos-de}) can do this job.

As we noted, the evolution of the interacting scenario at the background level is affected by presence of the interaction function $Q (t)$. Similarly, the evolution of the interacting scenario at the level of perturbations is also affected due to the presence of $Q (t)$.  Following the perturbed FLRW metric in any gauge \cite{Mukhanov:1990me,Ma:1995ey,Malik:2008im}, one can determine how the interacting scenario behaves in presence of the interaction function, see for instance Refs.~\cite{Valiviita:2008iv,Yang:2014gza,Wang:2014xca}.

\section{Different interacting scenarios and their roles in alleviating the $H_0$ tension}
\label{sec-models}

As already argued, the interaction function, $Q$, plays the key role in the interacting dynamics. However, as the choice of $Q$ is not unique, therefore, one can freely choose any interaction model and test its viability with the observational data.  
In the following we describe how the interaction models help in alleviating the Hubble constant tension. Throughout this section we have assumed that the curvature of the universe is zero.

\subsection{IDE with constant equation of state of DE}
\label{sec-IDE-const-w}

The interacting DE scenarios with constant equation of state in DE, $w_{de}$, have received tremendous attention in the community. 
In this context, the most simplest scenario is the one where the equation of state of DE mimics the cosmological constant, i.e. $w_{de} \sim -1$.  On the other hand, one can assume a constant $w_{de}$ other than $-1$ and investigate the consequences of the underlying interacting scenarios. In this section we shall discuss how different interacting models with constant $w_{de}$ 
return different values of $H_0$ and in all cases we consider the spatial flatness of the FLRW universe.

In Ref. \cite{Zhai:2023yny}, an interaction between DE and CDM  has been investigated using multiple CMB datasets from various experiments where the DE equation of state has been fixed at $w_{de} =-0.999$  and the interaction function has the following form 
\begin{eqnarray}\label{model-Q-const-w-rho-de}
    Q = H \xi \rho_{de}, 
\end{eqnarray}
where $\xi$ is the coupling parameter of the interaction function. Using different CMB data from multiple surveys, the constraints on $H_0$ are found to be as follows \cite{Zhai:2023yny}: for Planck 2018 alone, $H_0 = 71.6 \pm 2.1$ km/s/Mpc at 68\% CL and hence the tension with respect to R21 is reduced down to $0.6\sigma$; $H_0 = 72.6^{+3.4}_{-2.6}$ km/s/Mpc at 68\% CL (for ACT)\footnote{Here ACT refers to the full Atacama Cosmology Telescope temperature and polarization DR4 likelihood \cite{ACT:2020frw},
assuming a conservative Gaussian prior on $\tau = 0.065 \pm 0.015$ as in \cite{ACT:2020gnv}.} and hence the tension with respect to R21 is reduced down to $0.1\sigma$ which means in this case the $H_0$ tension is solved; for ACT+WMAP, we find $H_0 = 71.3 ^{+2.6}_{-3.2}$ km/s/Mpc at 68\% CL and hence the tension is reduced down to $0.6\sigma$; and for Planck 2018+ACT, $H_0 = 71.4^{+2.5}_{-2.8}$ km/s/Mpc at 68\% CL and in this case the tension is reduced down to $0.6\sigma$. We note that in all the cases, we find an evidence of $\xi \neq 0$ at more than 68\% CL for all these datasets \cite{Zhai:2023yny}.

In Ref. \cite{Yang:2019uog}, the authors investigated an interaction scenario between CDM and DE with the same interaction function  (\ref{model-Q-const-w-rho-de}), but $w_{de}$ was fixed at $-1$ which exactly mimics the vacuum energy sector. In this case, Planck 2018 alone leads to $H_ 0 =70.84^{+4.26}_{-2.50}$ km/s/Mpc at 68\% CL with an evidence of a non-zero coupling ($\xi =0.132^{+0.142}_{-0.077}$ at 68\% CL) at more than 68\% CL, and hence with respect to R21, the tension on $H_0$ is reduced down to $0.6\sigma$. This interacting scenario is quite appealing because even if in presence of massive neutrinos the tension in $H_0$ is significantly alleviated \cite{Yang:2019uog}.

In Ref. \cite{Yang:2020zuk}, another interacting scenario between CDM and DE with $w_{de} = -1$ has been investigated  for the  interaction function 

\begin{eqnarray}\label{model-Q-const-w-rho-de-2}
    Q = \Gamma \rho_{de}, 
\end{eqnarray}
which does not involve the Hubble parameter but it depends on the intrinsic nature of the dark sector, e.g. the energy density of the DE density. Here, $\Gamma$ is the coupling parameter of the interaction function having the dimension of the Hubble parameter and hence $\Gamma/H_0$ which is dimensionless.  For CMB alone from Planck 2018, this interaction model leads to $H_0 = 70.3^{+3.3}_{-2.0}$ km/s/Mpc at 68\% CL which thus alleviates the tension with R21 at $1\sigma$.  For Planck 2018+BAO, $H_0 = 69.0^{+1.4}_{-1.8}$ km/s/Mpc at 68\% CL, and as a consequence, the tension with R21 is reduced down to $2.1\sigma$.

In Ref. \cite{DiValentino:2019jae}, the authors considered the same interaction function (\ref{model-Q-const-w-rho-de}) describing an interaction between CDM and DE where $w_{de} \neq -1$ is a constant and it could lie either in the quintessence regime ($w_{de}> -1$) or in the phantom regime $(w_{de} < -1)$. 
For the quintessence case, Planck 2018 alone leads to $H_0 = 69.8^{+ 4}_{-2.5}$ km/s/Mpc at 68\% CL (hence, the tension with R21 is reduced down to $0.9 \sigma$) with an evidence of a non-zero coupling in the dark sector at more than $3\sigma$ significance \cite{DiValentino:2019jae}. When $w_{de}$ has a phantom nature, then the 68\% CL lower bound on the Hubble constant is found to be, $H_0 > 70.4$ km/s/Mpc with $w_{de} =-1.59^{+0.18}_{-0.33}$ (at 68\% CL for Planck 2018 alone). In this case the resolution of the $H_0$ tension appears due to the strong degeneracy between $H_0$-$w_{de}$, not due to the interaction between CDM and DE.

In Ref. \cite{Pan:2019jqh}, a special kind of an interaction scenario between CDM and DE (with constant $w_{de}$) in which the interaction function may change its sign during the evolution of the universe, known as sign changeable interaction function\footnote{For sign changeable interaction functions we refer to Refs. \cite{Wei:2010fz,Wei:2010cs,Li:2011ga,Guo:2017deu,Arevalo:2019axj}} was investigated for two variants of the interaction function. The interacting scenarios with $w_{de} \neq -1$ are found to increase the $H_0$ values and therefore here we shall report the constraints for this particular case. 
Considering the following interaction function

\begin{eqnarray}
    Q = 3 H \xi (\rho_{dm} - \rho_{de}), 
\end{eqnarray}
where $\xi$ is the coupling parameter of the interaction function,
it was found that Planck 2015+BAO returns $H_0 = 69.12^{+0.93}_{-1.39}$ km/s/Mpc at 68\% CL \cite{Pan:2019jqh} which thus alleviates the tension with R21 at $2.5\sigma$. 
For a generalized interaction function of this type 

\begin{eqnarray}
    Q = 3 H (\alpha \rho_{dm} - \beta \rho_{de}),
\end{eqnarray}
where $\alpha$, $\beta$ ($\alpha \neq \beta$) are the coupling parameters of this interaction function and they will have the same sign, Planck 2015+BAO gives $H_0 = 69.81^{+1.18}_{-1.39}$ km/s/Mpc at 68\% CL, and hence the tension with R21 is reduced down to $ 2\sigma$. The sign changeable interaction functions did not get much attention in the literature without any specific reason, however, they deserve further attention.

In Ref. \cite{Pan:2020bur}, an interacting scenario between CDM and DE with a constant $w_{de}$ other than $-1$, has been constrained with the following interaction 
function  

\begin{eqnarray}\label{model-Q-cons-w-trigonometric}
    Q =  3 H \xi \rho_{de} \sin\left(\frac{\rho_{de}}{\rho_{dm}}-1\right),
\end{eqnarray}
where $\xi$ is the coupling parameter of the interaction function. The analysis with Planck 2018 gives $H_0 = 72.67^{+5.43}_{-8.26}$ km/s/Mpc at 68\% CL and Planck 2018+BAO gives $H_0 = 69.17^{+1.52}_{-1.71}$
km/s/Mpc at 68\% CL \cite{Pan:2020bur}. Hence, for Planck 2018 and Planck 2018+BAO,  the tension on $H_0$ with respect to R21, is reduced down to $0.05 \sigma$, $2\sigma$, respectively. That means, for Planck 2018 alone case, the tension is completely solved. While it is important to mention that the resolution of the tension in this case is also influenced by the uncertainties.

In Ref. \cite{Pan:2022qrr},  an interacting scenario between a non-cold DM characterized by a non-null equation of state of DM, $w_{dm}~(\geq 0)$, and a DE component mimicking the vacuum energy sector, i.e. $w_{de} =-1$, has been explored for the interaction function $Q = 3 H \xi \rho_{de}$ of (\ref{model-Q-const-w-rho-de}). Unlike other interacting scenarios in which DM is assumed to be pressureless or cold, here, $w_{dm}$ is considered to be a free-to-vary parameter in $[0, 1]$ and the cosmological probes are allowed to decide its fate. This generalized interacting scenario is labeled as {\bf IWDM}.
We present the results in Table~\ref{tab:IVS} for different datasets. We notice a mild evidence of $w_{dm} \neq 0$ for all the datasets at 68\% CL, while at 95\% CL, this evidence goes away. In addition, an evidence of $\xi \neq 0$ is also  found from Planck 2018 and Planck 2018+Lensing datasets together with higher values of $H_0$ in both the datasets: $H_0 =   70.6_{-    2.4}^{+    4.3}$ km/s/Mpc at 68\% CL for Planck 2018 and $H_0 =    70.3_{-    2.8}^{+    4.3}$ km/s/Mpc at 68\% CL for Planck 2018+Lensing. Therefore, the tension in $H_0$ with respect to R21 is reduced down to $0.7 \sigma$ for both Planck 2018 and Planck 2018+Lensing.

\begin{table*}
\scalebox{0.65}
{                                                               
\begin{tabular}{ccccccccc}                                      
\hline\hline                                                                                                                    
Parameters & Planck 2018 & Planck 2018+Lensing & Planck 2018+BAO & Planck 2018+Pantheon & Planck 2018+Lensing+BAO+Pantheon\\ \hline

$\Omega_b h^2$ & $    0.02248_{-    0.00017-    0.00033}^{+    0.00017+    0.00035}$ & $    0.02249_{-    0.00016-    0.00032}^{+    0.00016+    0.00032}$  & $    0.02250_{-    0.00016-    0.00031}^{+    0.00016+    0.00032}$  & $    0.02247_{-    0.00017-    0.00032}^{+    0.00016+    0.00033}$ & $    0.02252_{-    0.00015-    0.00030}^{+    0.00015+    0.00031}$  \\

$\Omega_{dm} h^2$ & $    0.077_{- 0.056}^{+ 0.033} < 0.14$ & $    0.082_{-    0.036}^{+    0.055} < 0.14$  & $    0.104_{-    0.015-    0.040}^{+    0.023+    0.036}$  & $    0.109_{-    0.011-    0.024}^{+    0.013+    0.022}$  & $    0.112_{-    0.009-    0.020}^{+    0.011+    0.020}$  \\

$100\theta_{MC}$ & $    1.0436_{-    0.0036-    0.0046}^{+    0.0024+    0.0049}$ & $    1.0433_{-    0.0037-    0.0043}^{+    0.0020+    0.0051}$ & $    1.0419_{-    0.0014-    0.0021}^{+    0.0008+    0.0024}$  & $    1.04147_{-    0.00076-    0.0014}^{+    0.00065+    0.0015}$ & $    1.04137_{-    0.00066-    0.0012}^{+    0.00054+    0.0012}$ \\

$\tau$ & $    0.0534_{-    0.0074-    0.015}^{+    0.0075+    0.015}$ & $    0.0526_{-    0.0072-    0.014}^{+    0.0072+    0.015}$  & $    0.0541_{-    0.0075-    0.015}^{+    0.0077+    0.016}$   & $    0.0533_{-    0.0074-    0.015}^{+    0.0075+    0.015}$  & $    0.0539_{-    0.0071-    0.014}^{+    0.0070+    0.015}$ \\

$n_s$ & $    0.9628_{-    0.0046-    0.0088}^{+    0.0046+    0.0093}$ & $    0.9636_{-    0.0042-    0.0082}^{+    0.0042+    0.0082}$  & $    0.9637_{-    0.0042-    0.0083}^{+    0.0042+    0.0083}$   & $    0.9623_{-    0.0044-    0.0089}^{+    0.0044+    0.0088}$   & $    0.9646_{-    0.0039-    0.0078}^{+    0.0039+    0.0078}$   \\

${\rm{ln}}(10^{10} A_s)$ & $    3.046_{-    0.016-    0.030}^{+    0.015+    0.030}$ & $    3.043_{-    0.014-    0.028}^{+    0.014+    0.029}$ & $    3.046_{-    0.016-    0.031}^{+    0.016+    0.032}$   & $    3.046_{-    0.015-    0.030}^{+    0.015+    0.031}$  & $    3.044_{-    0.014-    0.027}^{+    0.014+    0.028}$  \\

$w_{dm}$ & $ 0.00122_{-    0.00097}^{+    0.00053} <0.0025$ & $    0.00112_{- 0.00091}^{+0.00047} < 0.0023$ & $    0.00115_{-    0.00098}^{+    0.00045} < 0.0025$  & $    0.00123_{-    0.00099}^{+    0.00052} < 0.0031$  & $    0.00108_{-    0.00096}^{+    0.00040} < 0.0023$    \\

$\xi$ & $    0.11_{-    0.07-    0.20}^{+    0.14+    0.17}$ & $    0.10_{-    0.08-    0.19}^{+    0.14+    0.17}$  & $    0.048_{-    0.064-    0.11}^{+    0.053+    0.12}$ & $    0.036_{-    0.036-    0.072}^{+    0.036+    0.072}$ & $    0.025_{-    0.032-    0.063}^{+    0.032+    0.063}$ \\

$\Omega_{m}$ & $    0.21_{-    0.14-    0.16}^{+    0.07+    0.19}$ & $    0.22_{-    0.14-    0.17}^{+    0.08+    0.18}$ & $    0.270_{-    0.046-    0.11}^{+    0.057+    0.10}$  & $    0.285_{-    0.032-    0.066}^{+    0.033+    0.066}$  & $    0.289_{-    0.028-    0.055}^{+    0.028+    0.055}$  \\

$H_0$ [Km/s/Mpc] & $   70.6_{-    2.4-    6.0}^{+    4.3+    5.2}$ & $   70.3_{-    2.8-    5.6}^{+    4.3+    5.4}$ & $   68.8_{-    1.6-    2.7}^{+    1.3+    3.0}$ & $   68.2_{-    1.0-    2.1}^{+    1.0+    2.1}$  & $   68.33_{-    0.81-    1.5}^{+    0.82+    1.6}$  \\

$S_8$ & $    1.01_{-    0.23-    0.27}^{+    0.08+    0.41}$ & $    0.98_{-    0.21-    0.24}^{+    0.06+    0.40}$ & $    0.880_{-    0.071-    0.11}^{+    0.032+    0.13}$  & $    0.869_{-    0.038-    0.066}^{+    0.028+    0.070}$ & $    0.849_{-    0.031-    0.052}^{+    0.022+    0.056}$ \\

$r_{\rm{drag}}$ [Mpc] & $  146.80_{-    0.33-    0.65}^{+    0.33+    0.64}$ & $  146.91_{-    0.29-    0.57}^{+    0.29+    0.56}$ & $  146.90_{-    0.31-    0.63}^{+    0.31+    0.61}$ & $  146.78_{-    0.33-    0.65}^{+    0.33+    0.64}$  & $  146.98_{-    0.27-    0.53}^{+    0.27+    0.53}$ \\

\hline\hline                                                                                                                    
\end{tabular} }                                                                
\caption{Summary of the observational constraints on the interacting scenario between a non-cold DM and vacuum considering various observational datasets.  Here, the free parameters of this interacting scenario are as follows: $\Omega_b h^2$ is the physical baryon density; $\Omega_{dm} h^2$ is the non-cold DM density; $\theta_{MC}$ is ratio of the sound horizon to the angular diameter distance; $\tau$ refers to the optical depth; $n_s$ is the spectral index; the amplitude of the primordial scalar perturbations is denoted by $A_s$. Among the derived parameters, $H_0$, $S_8$ are already defined; $\Omega_m$ refers to the density parameter of matter (non-cold DM+baryons) at present time and  $r_{\rm{drag}}$ refers to the comoving size of the sound horizon at $z_{\rm{drag}}$ (redshift at which baryon-drag optical depth equals unity). The table has been taken from  Ref. \cite{Pan:2022qrr}. }
\label{tab:IVS}                   
\end{table*}         
\begin{figure}
    \centering
    \includegraphics[width=1.1\textwidth]{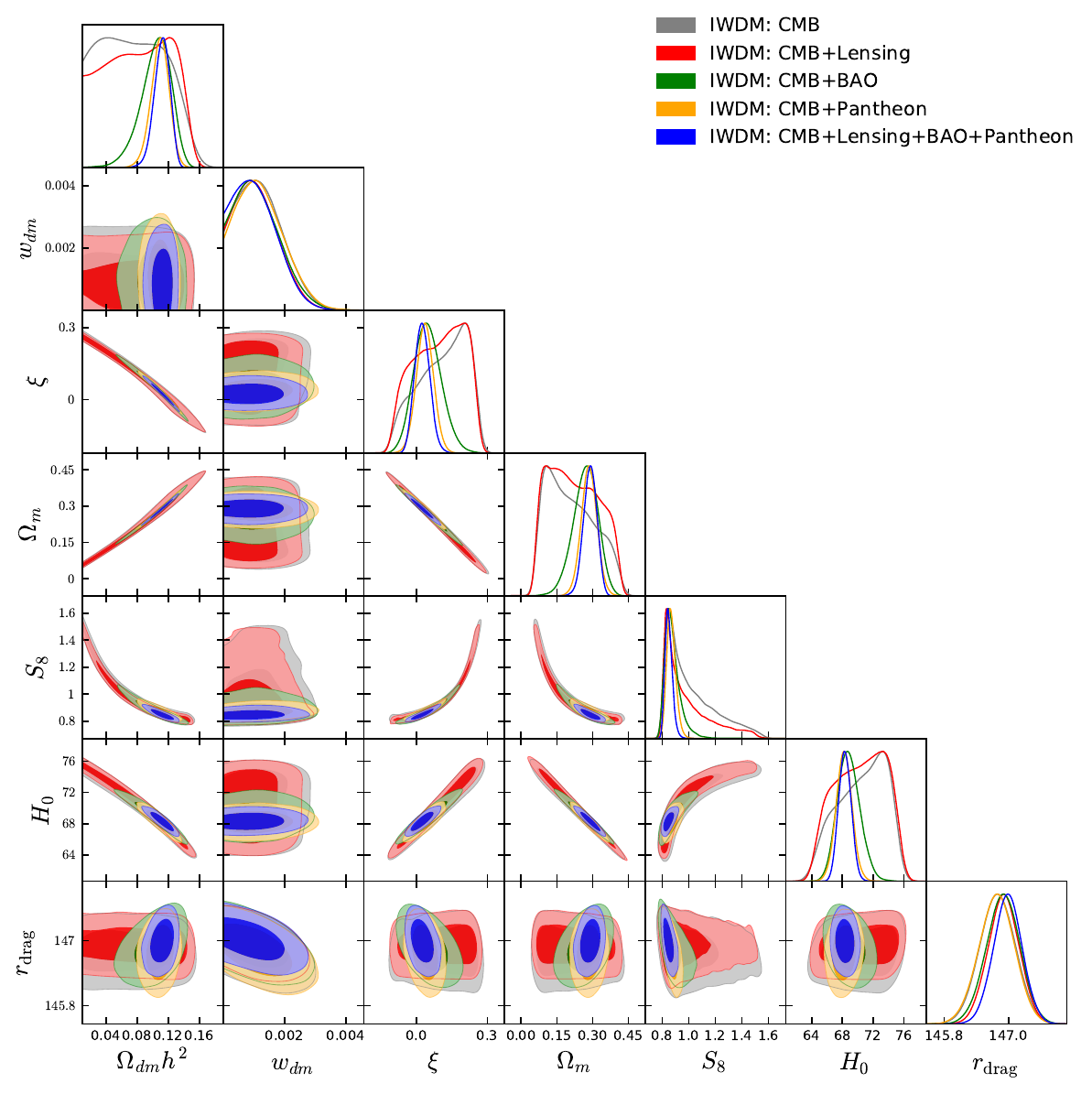}
    \caption{One dimensional marginalized posterior distributions of some of the free and derived parameters and the two dimensional joint contours for the interacting scenario between a non-cold DM and vacuum have been displayed considering various observational datasets.  Here ``CMB'' refers to Planck 2018 as described in \cite{Pan:2022qrr}. This figure has been taken from Ref. \cite{Pan:2022qrr}. }
    \label{fig:enter-label}
\end{figure}

\subsection{IDE with dynamical equation of state of DE}
\label{sec-IDE-variable-w}

The interacting scenarios in which the DE equation of state is dynamical, are the most generalized scenarios since such scenarios can recover the interacting scenarios with constant $w_{de}$ models as a special case. From both the theoretical and observational grounds, such scenarios are quite appealing. One can investigate such scenarios following two approaches $-$ either one can parametrize the DE equation of state in terms of some dynamical variable, e.g. the scale factor of the universe, see for instance Refs. \cite{Valiviita:2009nu,Majerotto:2009np,Pan:2016ngu,Sharov:2017iue,Yang:2017yme,Yang:2018uae,Pan:2019gop}, or, one can assume that DE is described by a scalar field $\phi$ with potential $V (\phi)$ where the equation of state of DE is dynamical being dependent on the scalar field and its potential, see Refs. \cite{Teixeira:2023zjt,Gomez-Valent:2022bku,Gomez-Valent:2020mqn,VanDeBruck:2017mua,vandeBruck:2016hpz}. Surprisingly, according to the existing literature, only a few works have been done considering the parametrized forms for $w_{de}$ \cite{Valiviita:2009nu,Majerotto:2009np,Pan:2016ngu,Sharov:2017iue,Yang:2017yme,Yang:2018uae,Pan:2019gop} without any particular reasons, while these scenarios and quite simple and they can lead to higher values of $H_0$. We suspect that one of the possible reasons could be the behaviour of the interacting scenarios at the level of perturbations which is certainly affected by the parametrized dynamical DE equation of state and an arbitrary choice of the dynamical $w_{de}$ may invite instabilities in the large scale. However, with a suitable choice of a dynamical $w_{de}$, one can cure the instability issues from the picture. 
In this section we shall discuss the impact of these scenarios in the context of the $H_0$ tension.

In Ref.~\cite{Pan:2019gop}, an interacting scenario between CDM and DE has been investigated with the following interaction function

\begin{eqnarray}\label{model-Q-variable-w-sec}
    Q = 3 H \xi (1+w_{de}) \rho_{de},
\end{eqnarray}
where $\xi$ is the coupling parameter of the interaction model and $w_{de}$ has a dynamical nature which can take one of the expressions below 

\begin{eqnarray}
&&{\rm Model~I}:~~w_{de}(a)=w_0\; a[1-\log(a)],\label{model1-dyn-w}\\
&&{\rm Model~II}:~~w_{de} (a) = w_0\; a \exp(1-a), \label{model2-dyn-w}\\
&&{\rm Model~III}:~~w_{de}(a)=w_0\; a[1+\sin(1-a)],\label{model3-dyn-w}\\
&&{\rm Model~IV}:~~w_{de}(a)=w_0 \; a[1+\arcsin(1-a)].\label{model4-dyn-w}
\end{eqnarray}
where $w_0$ in every model refers to the present value of $w_{de}$. In Table~\ref{tab-IDE-dynamical-w}, we present the estimated values of $H_0$ in each interacting  scenario for Planck 2015 and Planck 2015+BAO datasets~\cite{Pan:2019gop}. From Table~\ref{tab-IDE-dynamical-w}, it is clearly seen that for CMB data alone, the estimated values of $H_0$, for all dynamical equation of state parameters (\ref{model1-dyn-w}) $-$ (\ref{model4-dyn-w}), are significantly large compared to R21 by SH0ES collaboration \cite{Riess:2021jrx}, however, due to the large error bars, the tension with R21  is reduced down to less than $1\sigma$. 
However, for the Planck 2015+BAO dataset, the situation becomes quite interesting. As shown in Table~\ref{tab-IDE-dynamical-w}, the $H_0$ values are very close to its local estimation by SH0ES \cite{Riess:2021jrx} and the uncertainties on $H_0$ are significantly decreased compared yo Planck 2015. Thus, with respect to R21, the tension is reduced down to $1.1\sigma$ (for Model I), $0.7\sigma$ (for Model II), $0.3\sigma$ (for Model III), $0.1\sigma$ (for Model IV).  
We thus conclude that the last two models are quite sound since they can completely resolve the $H_0$ tension for both Planck 2018 and Planck 2015+BAO datasets.

\begin{table}[!ht]
\begin{center}
    \def\arraystretch{1.5}
	\begin{tabular}{llll}
    \hline\hline
		  Equation of state \quad\quad &\quad Hubble constant $H_0$ (km/s/Mpc) & \\
	&   Planck 2015 & Planck 2015+BAO \\ 
		\hline
		
		$w_{\rm DE} (a)=w_0a[1-\log(a)]$  \quad\quad & \quad  $ 81_{-14}^{+   13}$ & $   71.0_{-    1.5}^{+    1.5}$ \\
  
		$w_{\rm DE} (a)=w_0a\exp(1-a)$ \quad\quad & \quad  $ 84_{- 7}^{+   14} $ &  $   71.7_{-  1.7}^{+    1.5}$  \\
  
		$w_{\rm DE} (a)=w_0a[1+\sin(1-a)]$ \quad\quad & \quad $ 84_{-5}^{+ 12}$ & $   73.5_{- 1.7}^{+  1.6}$ \\
  
		$w_{\rm DE} (a)=w_0a[1+\arcsin(1-a)]$ \quad\quad & \quad $   82_{- 17}^{+   14}$ & $   72.8_{-    1.8}^{+    1.5}$   \\
		\hline
		\hline
	\end{tabular}
\end{center}
\caption{The table presents the constraints on $H_0$ for CMB and CMB+BAO datasets where CMB data have been taken from Planck 2015 considering various interacting scenarios in which the interaction function has the form (\ref{model-Q-variable-w-sec}) and the equation of state for DE is dynamical. }
	\label{tab-IDE-dynamical-w}
\end{table}

In the context of interacting scalar field models, the determination of the cosmological parameters is highly dependent on the choice of the scalar field potential and also on the interaction function as well.  As argued in Ref. \cite{Gomez-Valent:2020mqn}, such scalar field interacting models can also relieve the $H_0$ tension if a suitable choice of the potential is made. 

However, considering both the scalar field interacting scenarios and the parametrized $w_{de}$ interacting scenarios, 
we believe that both the scenarios are quite appealing in the light of the $H_0$ tension, and hence, they need further attention in the upcoming years.

\subsection{Interacting scenarios with time-dependent coupling parameter}
\label{sec-variable-coupling}

In most of the interaction functions explored in the literature, the coupling parameter is usually assumed to be constant. However, there is no such guiding principle available yet which dictates that the coupling parameter of the interaction functions should be constant instead of dynamical.  Even though the interaction scenarios with constant coupling parameter are relatively easier to handle, but
the interacting functions with time dependent coupling represent more generalized scenarios \cite{Li:2011ga,Guo:2017deu,Wang:2018azy,Yang:2019uzo,Yang:2020tax,Yao:2022kub,Yang:2022csz}. 
In the following we shall discuss that one can reconcile the Hubble constant tension within these generalized frameworks.

Let us focus on an interacting scenario between CDM and DE with constant equation-of-state $w_{de}$ and the interaction function has the following form 
\begin{eqnarray}
Q = 3 H \xi (\mathcal{X}) f (\rho_{dm}, \rho_{de}),\label{Q-variable-xi}
\end{eqnarray}
where $f$ is any analytic function of $\rho_{dm},~ \rho_{de}$; $\xi (\mathcal{X})$ is a dimensionless time-dependent coupling and $\mathcal{X}$ could be any cosmological variable.  We consider $\mathcal{X} = a/a_0 = a$ ($a_0$ is the present value of the scale factor and we set $a_0 =1$) and consider the first two terms of the Taylor series expansion of $\xi (a)$ around $a_0 = 1$ as follows 

\begin{eqnarray}\label{dynamical-coupling}
\xi (a) = \xi_0 + \xi_a \; (1-a)~,
\end{eqnarray}
where $\xi_0$, $\xi_a$ are constants and $\xi_0$ refers to the present day value of $\xi (a)$.  
Now, depending on the choice of $f (\rho_{dm}, \rho_{de})$, and the DE equation of state parameter, $w_{de}$, one could explore the dynamics of the underlying interacting scenario and estimate the constraints on the key cosmological parameters. 
We focus on the following two well known interaction functions of the form 
\begin{eqnarray}
&&Q = 3 \xi (\mathcal{X})\; H \; \rho_{de}, \label{variable-coupling-1}\\
&&Q= 3 \xi (\mathcal{X}) \; H\; \frac{\rho_{dm} \rho_{de}}{\rho_{dm}+ \rho_{de}},\label{variable-coupling-2}
\end{eqnarray}

\subsubsection{The case for $w_{de} = -1$}
\label{subsec-w=-1}

The simplest interacting scenario is the one when CDM interacts with the vacuum energy corresponding to  
the DE equation of state  $w_{de} = -1$, known as the Interacting Vacuum Scenario (IVS). 
The interacting scenarios driven by the interaction functions in eqn. (\ref{variable-coupling-1}) [labeled as IVS1],~ (\ref{variable-coupling-2}) [labeled as IVS2] together with the dynamical coupling $\xi (a)$ as in (\ref{dynamical-coupling}), were found to increase the Hubble constant value compared to the $\Lambda$CDM based Planck's estimation, see for instance, Ref. 
\cite{Yang:2019uzo}.  In particular, for the interaction function $Q = 3 H [\xi_0 + \xi_a (1-a)] \rho_{de}$ of eqn. (\ref{variable-coupling-1}), one finds \cite{Yang:2019uzo}
$H_0  =  69.2 \pm 2.8 $ km/sec/Mpc at 68\% CL (Planck 2015), and $H_0 = 68.5_{-    1.5}^{+    1.1}$ km/sec/Mpc at 68\% CL (Planck 2015+BAO) \cite{Yang:2019uzo}, and hence, with respect to R21, the tension is alleviated at $1.3\sigma$ and $2.7\sigma$, respectively.

On the other hand, for the interaction function $Q = 3 H [\xi_0 + \xi_a (1-a)] \frac{\rho_{dm}\rho_{de}}{\rho_{dm}+\rho_{de}}$ of eqn. (\ref{variable-coupling-2}),  one finds $H_0 = 68.3 \pm 3.5$ km/sec/Mpc at 68\% CL (Planck 2015) \cite{Yang:2019uzo} and $H_0 = 68.0_{-    1.5}^{+    1.3}$ km/sec/Mpc at 68\% CL (Planck 2015+BAO) \cite{Yang:2019uzo}, and hence with respect to R21, the tension is alleviated at $1.3 \sigma$ and $2.9 \sigma$, respectively. 

\begin{figure}
\includegraphics[width=0.49\textwidth]{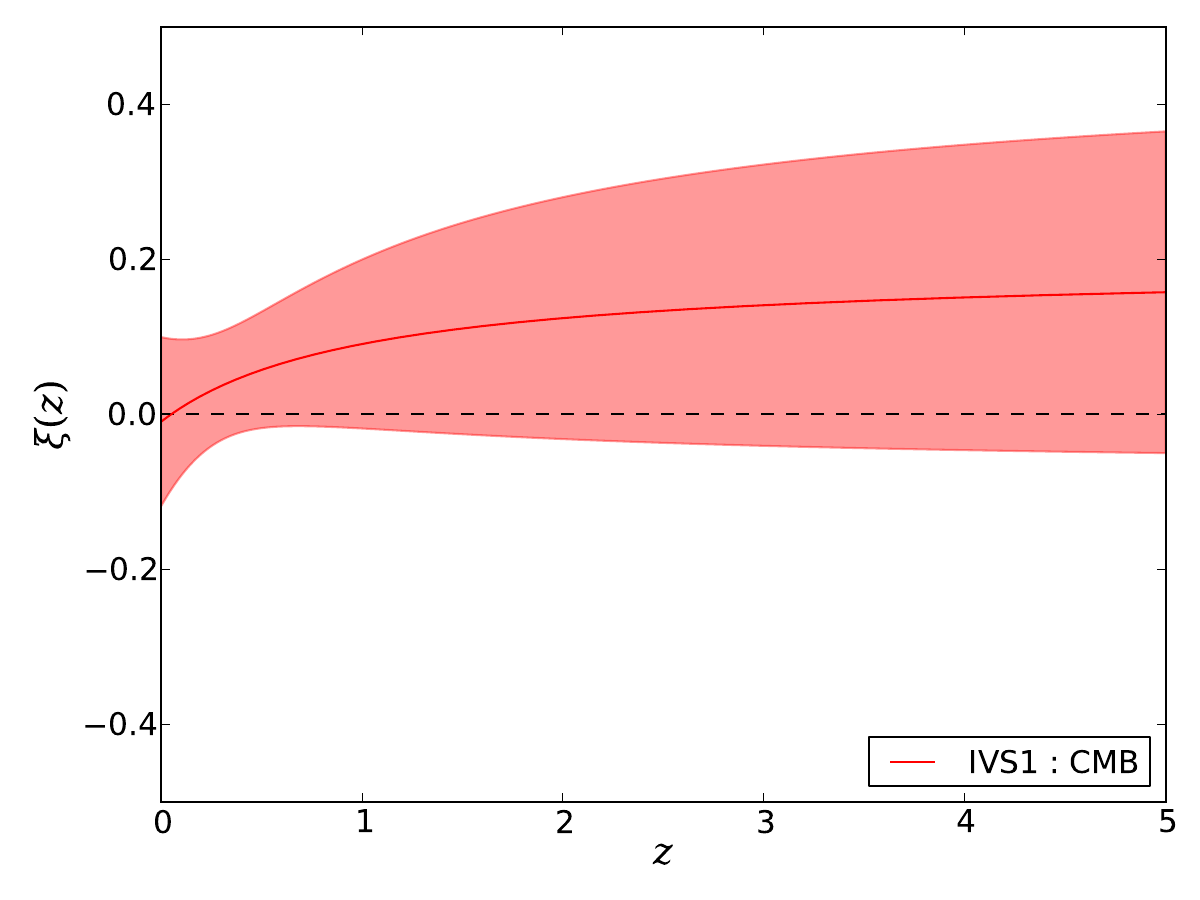}
\includegraphics[width=0.49\textwidth]{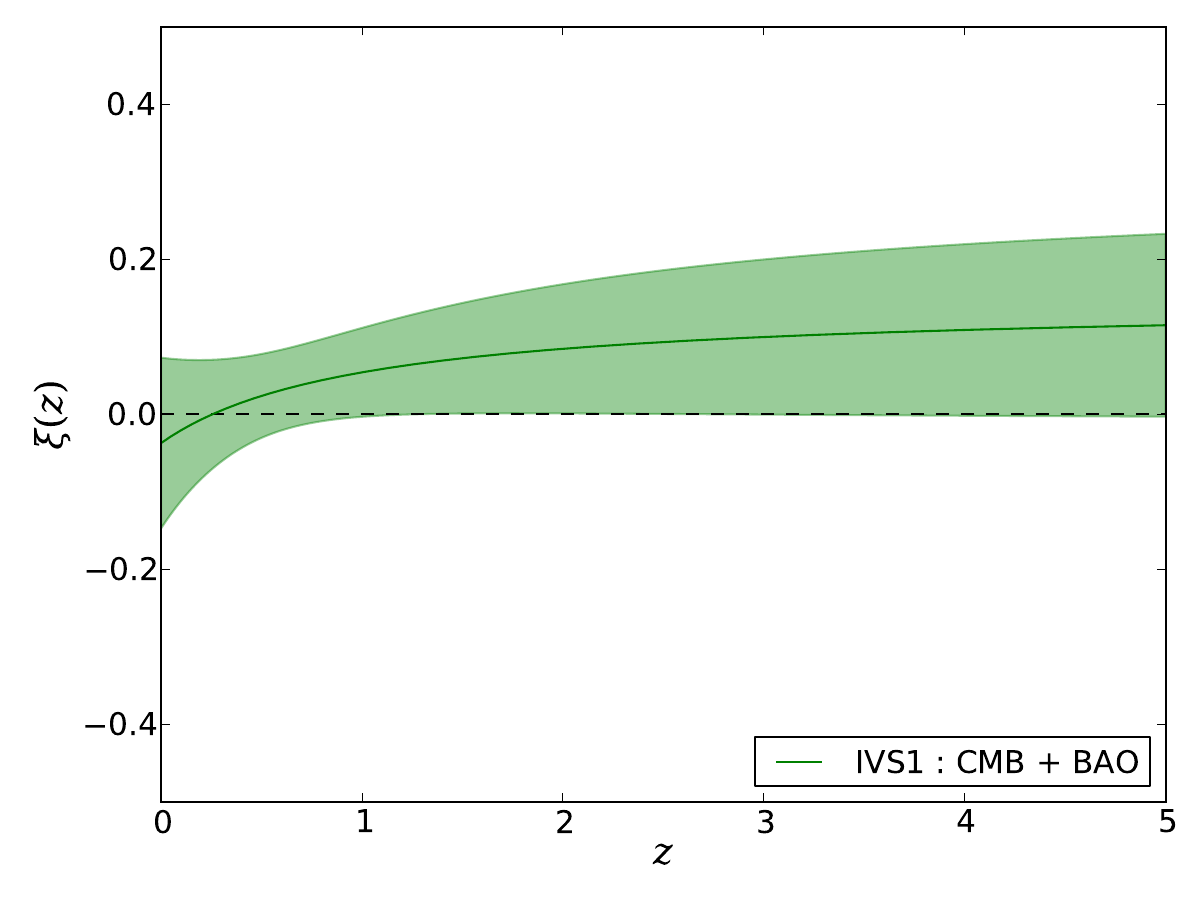}
    \caption{Evolution of the coupling parameter $\xi (z)$ [$1+z = a_0/a = 1/a$] of eqn. (\ref{dynamical-coupling}) for IVS1 for CMB (left graph) and CMB+BAO (right graph). The red (green) solid curve corresponds to the best fit curve for CMB (CMB+BAO) with the corresponding 68\% CL region.  
    Figure has been drawn taking the values of ($\xi_0, \xi_a$) from Table~III of 
    Ref. \cite{Yang:2019uzo}.}
    \label{fig:1-variable-coupling}
\end{figure}

\begin{figure}
\includegraphics[width=0.49\textwidth]{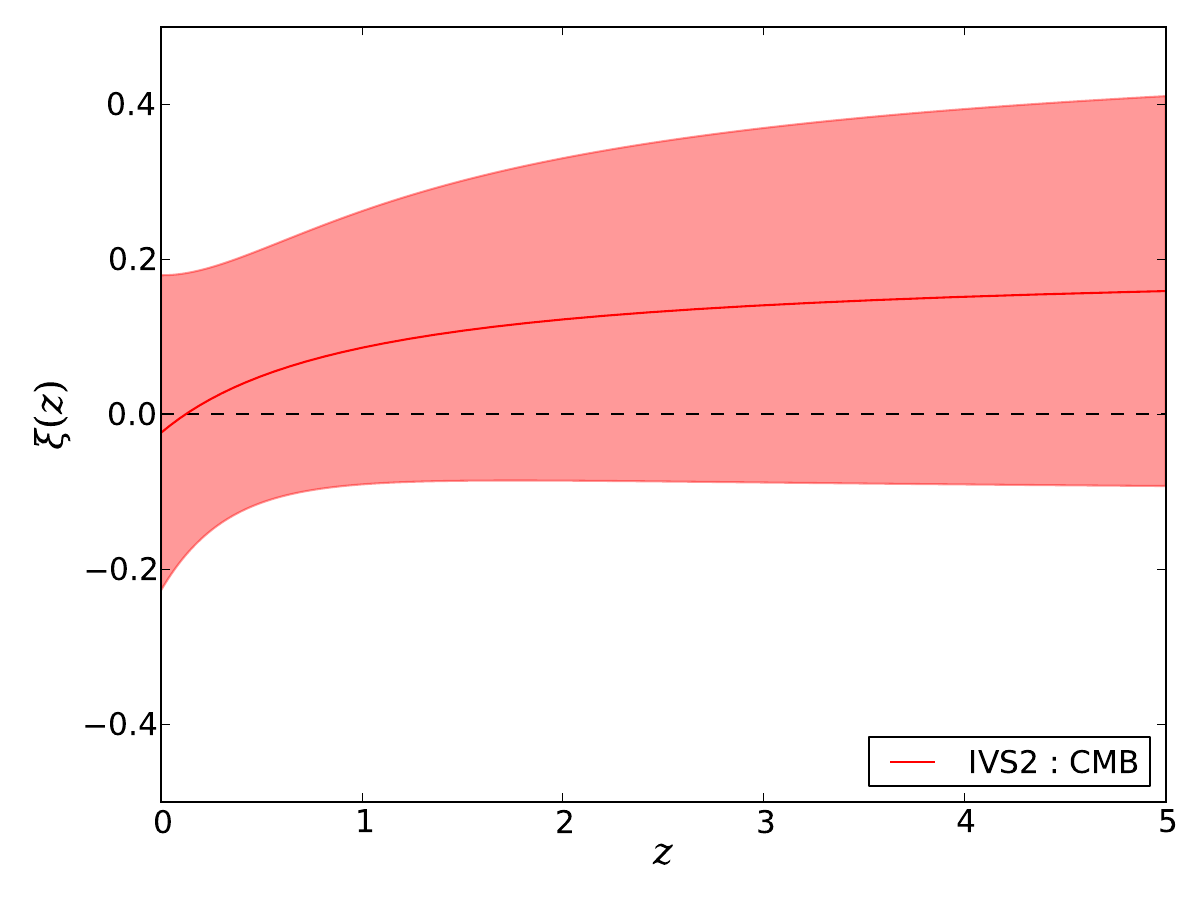}
\includegraphics[width=0.49\textwidth]{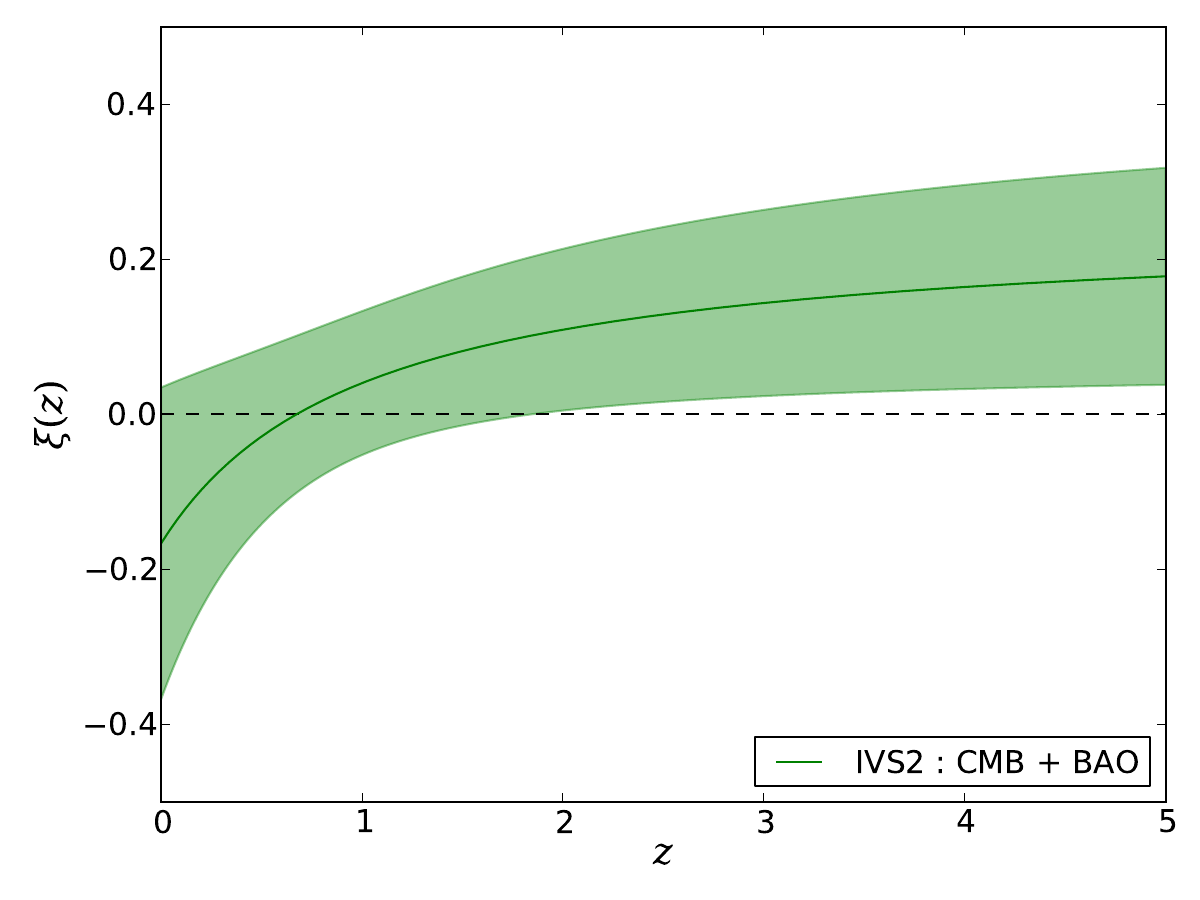}
    \caption{Evolution of the coupling parameter $\xi (z)$ [$1+z = a_0/a = 1/a$] of (\ref{dynamical-coupling}) for IVS2 for CMB (left graph) and CMB+BAO (right graph). The red (green) solid curve corresponds to the best fit curve for CMB (CMB+BAO) with the corresponding 68\% CL region.  
    Figure has been drawn taking the values of ($\xi_0, \xi_a$) from Table~III of Ref. \cite{Yang:2019uzo}. }
    \label{fig:2-variable-coupling}
\end{figure}

Now, referring to Table II and Table III of \cite{Yang:2019uzo}, one can find that even though the strong evidence for non-null values of $\xi_0$ and $\xi_a$ is not found from both Planck 2015 and Planck 2015+BAO, however, we cannot readily exclude the possibility of their non-null values. This can be evident from Fig. \ref{fig:1-variable-coupling} and Fig. \ref{fig:2-variable-coupling}
where we show the evolution of $\xi (z)$ [$1+z = a_0/a =1/a$] considering the 68\% CL values of $\xi_0$ and $\xi_a$ estimated from Planck 2015 and Planck 2015+BAO. One can notice from Fig. \ref{fig:1-variable-coupling} and Fig. \ref{fig:2-variable-coupling} that $\xi (z)$ has a transition from $\xi (z) > 0 $ to $\xi (z) <0$ which is prominent in the right graph of each Fig. \ref{fig:1-variable-coupling} and Fig. \ref{fig:2-variable-coupling}. This is an interesting observation which tells that for $\xi (z) > 0$
[equivalently, $Q (t) > 0$], the energy flow occurs from the cold DM to vacuum sector while for  $\xi (z) < 0$
[equivalently, $Q (t) < 0$] energy flow occurs from the vacuum energy sector to cold DM. Finally, 
as the choice of the dynamical coupling $\xi(a)$ is not unique, therefore, investigations with various choices for $\xi (a)$  will be certainly interesting. 


\subsubsection{The case for $w_{de} =$ constant $(\neq -1$)}

One can extend the interacting scenarios of the previous section \ref{subsec-w=-1} by a constant DE equation of state $w_{de}~(\neq -1)$ which is allowed to vary in a certain region.  Such extensions can lead to many interesting outcomes and also alleviate the Hubble constant tension \cite{Yang:2020tax}. As reported in Ref. \cite{Yang:2020tax}, if the DE equation of state $w_{de}$ is allowed to vary in the phantom regime (i.e. $w_{de} < -1$), then for the interaction function $Q = 3 H [\xi_0 + \xi_a (1-a)] \rho_{de}$, the $H_0$ is constrained to be $H_0 =  81.8_{- 11.7}^{+ 10.1}$ km/sec/Mpc at 68\% CL for Planck 2018 together with a mild evidence for $\xi_a = -0.077_{- 0.032}^{+ 0.064}$ at 68\% CL and a strong  preference for a phantom DE ($w_{de}-1.800_{- 0.387}^{+ 0.494}$  at 68\% CL), thus, the tension with R21 is alleviated within $1\sigma$. For Planck 2018+BAO dataset, the same scenario leads to $H_0 = 69.4_{- 1.9}^{+ 1.4}$ km/sec/Mpc at 68\% CL together with a strong preference for a phantom DE ($w_{de} = -1.207_{- 0.208}^{+ 0.196}$ at 95\% CL) but here the evidence for $\xi_a$ goes away. In this case, the tension with R21 is reduced down to $1.9 \sigma$.  On the other hand, for the quintessence DE ($w_{de}> -1$), the interaction function $Q = 3 H [\xi_0 + \xi_a (1-a)] \rho_{de}$ can lead to a higher value of the Hubble constant. For Planck 2018 alone, one finds that  $H_0 = 70.2_{- 3.1}^{+    4.1}$ km/sec/Mpc at 68\% CL, and thus the tension with R21 is resolved within $0.7\sigma$.  On the other hand, for Planck 2018+BAO, this interaction function leads to $H_0 = 68.4_{- 1.4}^{+ 1.3}$ km/sec/Mpc at 68\% CL which thus reduces the tension with R21 down to $2.7 \sigma$.

For the interaction function $Q = 3 H [\xi_0 + \xi_a (1-a)] \frac{\rho_{dm}\rho_{de}}{\rho_{dm}+\rho_{de}}$ of eqn. (\ref{variable-coupling-2}), if $w_{de}$ is allowed to vary in the phantom regime, then $H_0$ tension can be alleviated \cite{Yang:2020tax}. For Planck 2018 alone, this interacting scenario returns $H_0 > 72.5$ km/sec/Mpc at 95\% CL with a strong evidence of phantom DE ($w_{de} = -1.66652146_{- 0.370}^{+ 0.485}$ at 68\% CL) and for Planck 2018+BAO, $H_0 = 69.4_{-    1.5}^{+ 1.0}$ with a mild indication of the phantom DE ($w_{de} = -1.080_{-    0.031}^{+    0.067}$ at 68\% CL). Therefore, the tension in $H_0$ with R21 is reduced down to $2\sigma$ and $2.3\sigma$, respectively, for Planck 2018 alone and Planck 2018+BAO dataset.

\section{Interacting DE in a non-flat universe}
\label{sec-IDE+curvature}

In almost every works on the interacting scenarios, the curvature of our universe is usually assumed to be zero. This is usually motivated from some earlier investigations reporting a very small value of the curvature density parameter \cite{Gaztanaga:2008de,Mortonson:2009nw,Suyu:2013kha,LHuillier:2016mtc}, and, on the other hand, in presence of the curvature of the universe, the analysis of the underlying cosmological model becomes complicated a bit compared to the analysis of the models with no curvature. However,  with the growing sensitivity in the astronomical data, as the constraints on the cosmological parameters are getting improved, therefore,  there is no physical motivation to exclude the curvature parameter from the analysis. In fact, one can include the curvature parameter in the analysis of the cosmological models and finally allow the observational data to decide its fate. Some recent investigations argued that the observational evidences are in favour of a closed universe scenario \cite{DiValentino:2019qzk,Handley:2019tkm}. And it was also contended in \cite{DiValentino:2019qzk,Handley:2019tkm} that the preference of a closed universe may worsen the $H_0$ tension.
Thus, the question arises, can interacting scenarios in the context of a non-flat universe mitigate or solve the $H_0$ tension? This issue was investigated in Refs. \cite{DiValentino:2020kpf,Yang:2021hxg}.

In Ref. \cite{DiValentino:2020kpf}, an interaction between a CDM and DE with $w_{de} =-0.999$ was investigated for the interaction function (\ref{model-Q-const-w-rho-de}). As discussed in \cite{DiValentino:2020kpf}, different datasets return different values of $H_0$. For Planck 2018 alone, we find a lower value of $H_0$ with an evidence of a closed universe ($\Omega_k =-0.036^{+0.017}_{-0.013}$ at 68\% CL for Planck 2018) but no evidence of an interaction, however, from Planck 2018+BAO, while $\Omega_k$ is consistent to the flat universe, but we obtain an evidence of an interaction ($\xi = -0.32^{+0.31}_{-0.09}$ at 68\%) within 68\% CL together with 
$H_0 = 69.7 ^{+1.2}_{-1.6}$ km/s/Mpc at 68\% CL which thus alleviates the tension with R21 at $1.9\sigma$.

Further in Ref. \cite{Yang:2021hxg}, an interaction between the CDM and DE with $w_{de} \neq -1$ has been investigated with the interaction function $Q = 3H \xi \rho_{de}$.\footnote{Note that the interaction model (\ref{model-Q-const-w-rho-de}) is same with the interaction function $Q  = 3 H \xi \rho_{de}$. 
The presence of the factor $3$ has some mathematical importance, but no physics is associated with this choice. } 
For the large scale stability of the interacting scenario, the DE equation of state $w_{de}$ was divided into two separate regions, namely, $w_{de} < -1$ (phantom DE) and $w_{de}> -1$ (quintessence DE). We find that for $w_{de} < -1$, $H_0 = 69.0^{+1.2}_{-1.7}$ km/s/Mpc at 68\% CL for Planck 2018+BAO with an evidence of an interaction at 68\% CL ($\xi = -0.052^{-0.045}_{-0.025}$ at 68\% CL), however, the curvature parameter is consistent to a flat universe~\cite{Yang:2021hxg}. Therefore, in this case, the tension with R21 is alleviated at $2.3 \sigma$ for Planck 2018+BAO.  For $w_{de} > -1$, one leads to $H_0 = 68.5 \pm 1.4$ km/s/Mpc at 68\% CL for Planck 2018+BAO \cite{Yang:2021hxg}, and hence, the tension with R21 is reduced down to $2.6 \sigma$. 

The interaction functions presented in this section are not the only choices. One can investigate other kind of interaction functions considering the fact that $w_{de}$ could be dynamical. We anticipate that the interacting scenarios in the context of a non-flat universe are promising and these models need further investigations with the upcoming observational data from various astronomical surveys.

\section{Summary and Conclusions}
\label{sec-summary}

The theme of the present chapter is the theory of non-gravitational interaction between two heavy dark components of the universe, namely, DM and DE. The interaction between them is characterized by the transfer or energy between them. Over the last couple of years, interacting DE-DM scenarios have gained remarkable attention in  the scientific community for many appealing consequences, such as, alleviation of the cosmic coincidence problem, crossing of the phantom divide line, and addressing the cosmological tensions. The heart of this theory is the interaction function $Q (t)$ that governs the transfer of energy between the dark sectors, and, as a consequence, this interaction function modifies the evolution equations for DE and DM at the level of background and perturbations. Thus, it is natural to expect that the cosmological parameters should be affected due to the presence of this interaction function. However, there is no such fundamental principle available yet for deriving the interaction function. Even though some attempts have been made to formulate the interaction function,  but no conclusive statement has been made so far in this direction. 
Nevertheless, due to the existing features available in this particular theory, interacting DE has been proved as a phenomenologically rich cosmological theory.

In this chapter we have discussed a variety of interacting scenarios between DE and DM characterized by the triplet $(Q(t),~ w_{dm},~ w_{de})$ where $Q (t)$ has a wide variations; $w_{dm}$, representing the equation of state of the DM sector could be either zero or some non-negative constant, however, the DE equation of state, $w_{de}$, could be either constant or time dependent. 
Initially we have focused on the interacting scenarios assuming that the curvature of the universe is zero. In this set up,   we have found that depending on the nature of $w_{de}$ and the interaction function $Q (t)$, the $5\sigma$ tension on $H_0$ between Planck  \cite{Planck:2018vyg} and SH0ES collaboration \cite{Riess:2021jrx} can be significantly reduced. In particular, in some models, $H_0$ tension is reduced down to $ < 3\sigma$; in some scenarios this tension is reduced down to $< 2\sigma$; in some scenarios it is reduced down to $< 1 \sigma$; and in some scenarios, one can completely solve this tension.   On the other hand, assuming that our universe may not be spatially flat, we have discussed the impact of the interacting scenarios on the $H_0$ tension.

Based on the discussions presented in this chapter, it is pretty clear that the interacting scenarios between DE and DM are really promising for reconciling the Hubble constant tension.  In addition, we would like to stress that there are many gaps in this domain which need to be explored crucially with the upcoming cosmological probes from several astronomical surveys.


\begin{acknowledgement}
SP acknowledges the financial support from  the Department of Science and Technology (DST), Govt. of India under the Scheme  ``Fund for Improvement of S\&T Infrastructure (FIST)'' [File No. SR/FST/MS-I/2019/41].  WY acknowledges the  support by the National Natural Science Foundation of China under Grants No. 12175096 and No. 11705079, and Liaoning Revitalization Talents Program under Grant no. XLYC1907098.
\end{acknowledgement}





\begin{thebibliography}{99}

\bibitem{SupernovaSearchTeam:1998fmf}
A.~G.~Riess \textit{et al.} [Supernova Search Team],
Astron. J. \textbf{116}, 1009-1038 (1998)
[arXiv:astro-ph/9805201 [astro-ph]].

\bibitem{SupernovaCosmologyProject:1998vns}
S.~Perlmutter \textit{et al.} [Supernova Cosmology Project],
Astrophys. J. \textbf{517}, 565-586 (1999)
[arXiv:astro-ph/9812133 [astro-ph]].


\bibitem{Weinberg:1988cp}
S.~Weinberg,
Rev. Mod. Phys. \textbf{61}, 1-23 (1989)

\bibitem{Zlatev:1998tr}
I.~Zlatev, L.~M.~Wang and P.~J.~Steinhardt,
Phys. Rev. Lett. \textbf{82}, 896-899 (1999)
[arXiv:astro-ph/9807002 [astro-ph]].

\bibitem{Planck:2018vyg}
N.~Aghanim \textit{et al.} [Planck],
Astron. Astrophys. \textbf{641}, A6 (2020)
[erratum: Astron. Astrophys. \textbf{652}, C4 (2021)]
[arXiv:1807.06209 [astro-ph.CO]].

\bibitem{Riess:2021jrx}
A.~G.~Riess, W.~Yuan, L.~M.~Macri, D.~Scolnic, D.~Brout, S.~Casertano, D.~O.~Jones, Y.~Murakami, L.~Breuval and T.~G.~Brink, \textit{et al.}
Astrophys. J. Lett. \textbf{934}, no.1, L7 (2022)
[arXiv:2112.04510 [astro-ph.CO]].

\bibitem{Riess:2022mme}
A.~G.~Riess, L.~Breuval, W.~Yuan, S.~Casertano, L.~M.~Macri, J.~B.~Bowers, D.~Scolnic, T.~Cantat-Gaudin, R.~I.~Anderson and M.~C.~Reyes,
Astrophys. J. \textbf{938}, no.1, 36 (2022)
[arXiv:2208.01045 [astro-ph.CO]].

\bibitem{Asgari:2019fkq}
M.~Asgari, T.~Tr\"oster, C.~Heymans, H.~Hildebrandt, J.~L.~van den Busch, A.~H.~Wright, A.~Choi, T.~Erben, B.~Joachimi and S.~Joudaki, \textit{et al.}
Astron. Astrophys. \textbf{634}, A127 (2020)
[arXiv:1910.05336 [astro-ph.CO]].

\bibitem{KiDS:2020suj}
M.~Asgari \textit{et al.} [KiDS],
Astron. Astrophys. \textbf{645}, A104 (2021)
[arXiv:2007.15633 [astro-ph.CO]].

\bibitem{Joudaki:2019pmv}
S.~Joudaki, H.~Hildebrandt, D.~Traykova, N.~E.~Chisari, C.~Heymans, A.~Kannawadi, K.~Kuijken, A.~H.~Wright, M.~Asgari and T.~Erben, \textit{et al.}
Astron. Astrophys. \textbf{638}, L1 (2020)
[arXiv:1906.09262 [astro-ph.CO]].

\bibitem{DES:2021wwk}
T.~M.~C.~Abbott \textit{et al.} [DES],
Phys. Rev. D \textbf{105}, no.2, 023520 (2022)
[arXiv:2105.13549 [astro-ph.CO]].

\bibitem{DES:2021bvc}
A.~Amon \textit{et al.} [DES],
Phys. Rev. D \textbf{105}, no.2, 023514 (2022)
[arXiv:2105.13543 [astro-ph.CO]].

\bibitem{DES:2021vln}
L.~F.~Secco \textit{et al.} [DES],
Phys. Rev. D \textbf{105}, no.2, 023515 (2022)
[arXiv:2105.13544 [astro-ph.CO]].

\bibitem{KiDS:2021opn}
A.~Loureiro \textit{et al.} [KiDS and Euclid],
Astron. Astrophys. \textbf{665}, A56 (2022)
[arXiv:2110.06947 [astro-ph.CO]].

\bibitem{Heymans:2020gsg}
C.~Heymans, T.~Tr\"oster, M.~Asgari, C.~Blake, H.~Hildebrandt, B.~Joachimi, K.~Kuijken, C.~A.~Lin, A.~G.~S\'anchez and J.~L.~van den Busch, \textit{et al.}
Astron. Astrophys. \textbf{646}, A140 (2021)
[arXiv:2007.15632 [astro-ph.CO]].

\bibitem{DES:2020ahh}
T.~M.~C.~Abbott \textit{et al.} [DES],
Phys. Rev. D \textbf{102}, no.2, 023509 (2020)
[arXiv:2002.11124 [astro-ph.CO]].

\bibitem{Philcox:2021kcw}
O.~H.~E.~Philcox and M.~M.~Ivanov,
Phys. Rev. D \textbf{105}, no.4, 043517 (2022)
[arXiv:2112.04515 [astro-ph.CO]].


\bibitem{DiValentino:2021izs}
E.~Di Valentino, O.~Mena, S.~Pan, L.~Visinelli, W.~Yang, A.~Melchiorri, D.~F.~Mota, A.~G.~Riess and J.~Silk,
Class. Quant. Grav. \textbf{38}, no.15, 153001 (2021)
[arXiv:2103.01183 [astro-ph.CO]].

\bibitem{Perivolaropoulos:2021jda}
L.~Perivolaropoulos and F.~Skara,
New Astron. Rev. \textbf{95}, 101659 (2022)
[arXiv:2105.05208 [astro-ph.CO]].

\bibitem{Abdalla:2022yfr}
E.~Abdalla, G.~Franco Abell\'an, A.~Aboubrahim, A.~Agnello, O.~Akarsu, Y.~Akrami, G.~Alestas, D.~Aloni, L.~Amendola and L.~A.~Anchordoqui, \textit{et al.}
JHEAp \textbf{34}, 49-211 (2022)
[arXiv:2203.06142 [astro-ph.CO]].

\bibitem{Kamionkowski:2022pkx}
M.~Kamionkowski and A.~G.~Riess,
[arXiv:2211.04492 [astro-ph.CO]].


\bibitem{Wetterich:1994bg}
C.~Wetterich,
Astron. Astrophys. \textbf{301}, 321-328 (1995)
[arXiv:hep-th/9408025 [hep-th]].

\bibitem{Amendola:1999er} 
  L.~Amendola,
  {\it Coupled quintessence,}
  Phys.\ Rev.\ D {\bf 62}, 043511 (2000)
  [astro-ph/9908023].


 \bibitem{Chimento:2003iea}
L.~P.~Chimento, A.~S.~Jakubi, D.~Pavon and W.~Zimdahl,
Phys. Rev. D \textbf{67}, 083513 (2003)
[arXiv:astro-ph/0303145 [astro-ph]]. 
  
  \bibitem{Cai:2004dk} 
  R.~G.~Cai and A.~Wang,
  JCAP {\bf 0503}, 002 (2005)
  [hep-th/0411025].

\bibitem{Pavon:2005yx} 
D.~Pav\'{o}n and W.~Zimdahl, 
Phys.\ Lett.\ B {\bf 628}, 206 (2005)
[arXiv:gr-qc/0505020]. 

\bibitem{delCampo:2006vv}
S.~del Campo, R.~Herrera, G.~Olivares and D.~Pavon,
Phys. Rev. D \textbf{74}, 023501 (2006)
[arXiv:astro-ph/0606520 [astro-ph]].

\bibitem{Berger:2006db}
M.~S.~Berger and H.~Shojaei,
Phys. Rev. D \textbf{73}, 083528 (2006)
[arXiv:gr-qc/0601086 [gr-qc]]. 


 \bibitem{delCampo:2008sr} 
S.~del Campo, R.~Herrera and D.~Pav\'{o}n, 
Phys.\ Rev.\ D {\bf 78}, 021302 (2008)
[arXiv:0806.2116 [astro-ph]].



\bibitem{delCampo:2008jx} 
S.~del Campo, R.~Herrera and D.~Pav\'{o}n, 
JCAP {\bf 0901}, 020 (2009)
[arXiv:0812.2210 [gr-qc]].

 
 \bibitem{Barrow:2006hia} 
  J.~D.~Barrow and T.~Clifton,
  Phys.\ Rev.\ D {\bf 73}, 103520 (2006)
  [gr-qc/0604063].
  
 
  \bibitem{Amendola:2006dg}
  L.~Amendola, G.~Camargo Campos and R.~Rosenfeld,
  Phys.\ Rev.\ D {\bf 75}, 083506 (2007)
  [astro-ph/0610806].

 \bibitem{Koivisto:2006ai} 
  T.~Koivisto and D.~F.~Mota,
  Phys.\ Rev.\ D {\bf 75}, 023518 (2007)
  [hep-th/0609155].
  
\bibitem{He:2008tn}
  J.~H.~He and B.~Wang,
  JCAP {\bf 0806}, 010 (2008)
  [arXiv:0801.4233 [astro-ph]].
  
  
  
 
  
 \bibitem{Gavela:2009cy} 
  M.~B.~Gavela, D.~Hernandez, L.~Lopez Honorez, O.~Mena and S.~Rigolin,
  JCAP {\bf 0907}, 034 (2009)
  [arXiv:0901.1611 [astro-ph.CO]].




  
  \bibitem{Majerotto:2009np} 
  E.~Majerotto, J.~V\"{a}liviita and R.~Maartens,
  Mon.\ Not.\ Roy.\ Astron.\ Soc.\  {\bf 402}, 2344 (2010)
  [arXiv:0907.4981 [astro-ph.CO]].


    \bibitem{Valiviita:2009nu}
J.~Valiviita, R.~Maartens and E.~Majerotto,
Mon. Not. Roy. Astron. Soc. \textbf{402}, 2355-2368 (2010)
[arXiv:0907.4987 [astro-ph.CO]].
  
   \bibitem{Clemson:2011an} 
  T.~Clemson, K.~Koyama, G.~B.~Zhao, R.~Maartens and J.~V\"{a}liviita,
  Phys.\ Rev.\ D {\bf 85}, 043007 (2012)
  [arXiv:1109.6234 [astro-ph.CO]].
  
  \bibitem{Salvatelli:2014zta} 
  V.~Salvatelli, N.~Said, M.~Bruni, A.~Melchiorri and D.~Wands,
  Phys.\ Rev.\ Lett.\  {\bf 113}, no. 18, 181301 (2014)
  [arXiv:1406.7297 [astro-ph.CO]].
  
  
\bibitem{Yang:2014vza}
  W.~Yang and L.~Xu,
  JCAP {\bf 1408}, 034 (2014)
  [arXiv:1401.5177 [astro-ph.CO]].
  
  \bibitem{Yang:2014gza}  
  W.~Yang and L.~Xu,
  Phys.\ Rev.\ D {\bf 89}, no.8,  083517 (2014)
  [arXiv:1401.1286 [astro-ph.CO]].
  
  \bibitem{Wang:2014xca}
Y.~Wang, D.~Wands, G.~B.~Zhao and L.~Xu,
Phys. Rev. D \textbf{90}, no.2, 023502 (2014)
doi:10.1103/PhysRevD.90.023502
[arXiv:1404.5706 [astro-ph.CO]].
  
  \bibitem{Faraoni:2014vra} 
  V.~Faraoni, J.~B.~Dent and E.~N.~Saridakis,
  Phys.\ Rev.\ D {\bf 90}, no. 6, 063510 (2014)
  [arXiv:1405.7288 [gr-qc]].
  
  \bibitem{Pan:2012ki}
  S.~Pan, S.~Bhattacharya and S.~Chakraborty,
  Mon.\ Not.\ Roy.\ Astron.\ Soc.\  {\bf 452}, no.3,  3038 (2015)
  [arXiv:1210.0396 [gr-qc]].
  
 
  
  \bibitem{Yang:2014hea} 
  W.~Yang and L.~Xu,
  Phys.\ Rev.\ D {\bf 90}, no. 8, 083532 (2014)
  [arXiv:1409.5533 [astro-ph.CO]]. 
  
  
 
  
  \bibitem{Li:2015vla} 
  Y.~H.~Li, J.~F.~Zhang and X.~Zhang,
  Phys.\ Rev.\ D {\bf 93}, no. 2, 023002 (2016)
  [arXiv:1506.06349 [astro-ph.CO]].
  
  
  
  \bibitem{Nunes:2016dlj} 
  R.~C.~Nunes, S.~Pan and E.~N.~Saridakis,
  Phys.\ Rev.\ D {\bf 94}, no. 2, 023508 (2016)
  [arXiv:1605.01712 [astro-ph.CO]].
  
  \bibitem{Yang:2016evp}
  W.~Yang, H.~Li, Y.~Wu and J.~Lu,
  JCAP {\bf 1610}, no.10,  007 (2016)
  [arXiv:1608.07039 [astro-ph.CO]].
  
  \bibitem{Pan:2016ngu} 
  S.~Pan and G.~S.~Sharov,
  Mon.\ Not.\ Roy.\ Astron.\ Soc.\  {\bf 472}, no. 4, 4736 (2017)
  [arXiv:1609.02287 [gr-qc]].
  
  \bibitem{Mukherjee:2016shl} 
  A.~Mukherjee and N.~Banerjee,
  Class.\ Quant.\ Grav.\  {\bf 34}, no. 3, 035016 (2017)
  [arXiv:1610.04419 [astro-ph.CO]].
 
  \bibitem{Sharov:2017iue} 
  G.~S.~Sharov, S.~Bhattacharya, S.~Pan, R.~C.~Nunes and S.~Chakraborty,
  Mon.\ Not.\ Roy.\ Astron.\ Soc.\  {\bf 466}, no. 3, 3497 (2017)
  [arXiv:1701.00780 [gr-qc]].
  
  
\bibitem{Guo:2017hea} 
  R.~Y.~Guo, Y.~H.~Li, J.~F.~Zhang and X.~Zhang,
  JCAP {\bf 1705}, no. 05, 040 (2017)
  [arXiv:1702.04189 [astro-ph.CO]].


 \bibitem{Cai:2017yww} 
  R.~G.~Cai, N.~Tamanini and T.~Yang,
  JCAP {\bf 1705}, no. 05, 031 (2017)
  [arXiv:1703.07323 [astro-ph.CO]].
  
  
  
  \bibitem{Yang:2017yme} 
  W.~Yang, N.~Banerjee and S.~Pan,
  Phys.\ Rev.\ D {\bf 95}, no. 12, 123527 (2017)
  [arXiv:1705.09278 [astro-ph.CO]].
  
 \bibitem{Yang:2017ccc} 
  W.~Yang, S.~Pan and D.~F.~Mota,
  Phys.\ Rev.\ D {\bf 96}, no. 12, 123508 (2017)
  [arXiv:1709.00006 [astro-ph.CO]]. 
  
  \bibitem{Yang:2017zjs} 
  W.~Yang, S.~Pan and J.~D.~Barrow,
  Phys.\ Rev.\ D {\bf 97}, no. 4, 043529 (2018)
  [arXiv:1706.04953 [astro-ph.CO]].
 
  \bibitem{Pan:2017ent} 
  S.~Pan, A.~Mukherjee and N.~Banerjee,
  Mon.\ Not.\ Roy.\ Astron.\ Soc.\  {\bf 477}, no. 1, 1189 (2018)
  [arXiv:1710.03725 [astro-ph.CO]]. 
  
  
  \bibitem{Yang:2018pej} 
  W.~Yang, S.~Pan and A.~Paliathanasis,
  Mon.\ Not.\ Roy.\ Astron.\ Soc.\  {\bf 482}, no. 1, 1007 (2019)
  [arXiv:1804.08558 [gr-qc]].
  
   \bibitem{Yang:2018ubt} 
  W.~Yang, S.~Pan, L.~Xu and D.~F.~Mota,
  Mon.\ Not.\ Roy.\ Astron.\ Soc.\  {\bf 482}, no. 2, 1858 (2019)
  [arXiv:1804.08455 [astro-ph.CO]].
  
  
  \bibitem{Paliathanasis:2019hbi} 
  A.~Paliathanasis, S.~Pan and W.~Yang,
Int. J. Mod. Phys. D \textbf{28}, no. 12, 1950161 (2019)
  arXiv:1903.02370 [gr-qc].
  
  
 
  \bibitem{Pan:2019jqh} 
  S.~Pan, W.~Yang, C.~Singha and E.~N.~Saridakis,
Phys. Rev. D \textbf{100}, no.8, 083539 (2019)
  arXiv:1903.10969 [astro-ph.CO].
  
  
  \bibitem{Yang:2019bpr} 
  W.~Yang, S.~Pan, E.~Di Valentino, B.~Wang and A.~Wang,
  JCAP \textbf{05}, 050 (2020)
  arXiv:1904.11980 [astro-ph.CO].
  
  
  
  \bibitem{Yang:2019vni} 
  W.~Yang, S.~Vagnozzi, E.~Di Valentino, R.~C.~Nunes, S.~Pan and D.~F.~Mota,
  JCAP {\bf 1907}, 037 (2019)
  [arXiv:1905.08286 [astro-ph.CO]].
  
  

\bibitem{Yang:2019uog}
W.~Yang, S.~Pan, R.~C.~Nunes and D.~F.~Mota,
JCAP \textbf{04}, 008 (2020)
doi:10.1088/1475-7516/2020/04/008
[arXiv:1910.08821 [astro-ph.CO]].

\bibitem{Yang:2020zuk}
W.~Yang, E.~Di Valentino, S.~Pan, S.~Basilakos and A.~Paliathanasis,
Phys. Rev. D \textbf{102}, no.6, 063503 (2020)
doi:10.1103/PhysRevD.102.063503
[arXiv:2001.04307 [astro-ph.CO]].

\bibitem{Pan:2020bur}
S.~Pan, W.~Yang and A.~Paliathanasis,
Mon. Not. Roy. Astron. Soc. \textbf{493}, no.3, 3114-3131 (2020)
doi:10.1093/mnras/staa213
[arXiv:2002.03408 [astro-ph.CO]].




  
 
  \bibitem{Bolotin:2013jpa} 
  Y.~L.~Bolotin, A.~Kostenko, O.~A.~Lemets and D.~A.~Yerokhin,
  Int.\ J.\ Mod.\ Phys.\ D {\bf 24}, no. 03, 1530007 (2015)
  [arXiv:1310.0085 [astro-ph.CO]].
  
  \bibitem{Wang:2016lxa}
  B.~Wang, E.~Abdalla, F.~Atrio-Barandela and D.~Pav\'{o}n,
  Rept.\ Prog.\ Phys.\  {\bf 79} (2016) no.9,  096901
  [arXiv:1603.08299 [astro-ph.CO]].
 

\bibitem{Wang:2005jx} 
  B.~Wang, Y.~g.~Gong and E.~Abdalla,
  Phys.\ Lett.\ B {\bf 624}, 141 (2005)
  [hep-th/0506069].

\bibitem{Das:2005yj} 
  S.~Das, P.~S.~Corasaniti and J.~Khoury,
  Phys.\ Rev.\ D {\bf 73}, 083509 (2006)
  [astro-ph/0510628].
  
  
  \bibitem{Sadjadi:2006qb} 
  H.~M.~Sadjadi and M.~Honardoost,
  Phys.\ Lett.\ B {\bf 647}, 231 (2007)
  [gr-qc/0609076].


\bibitem{Pan:2014afa} 
  S.~Pan and S.~Chakraborty,
  Int.\ J.\ Mod.\ Phys.\ D {\bf 23}, no. 11, 1450092 (2014)
  [arXiv:1410.8281 [gr-qc]].
  
 
 
  \bibitem{Kumar:2016zpg} 
  S.~Kumar and R.~C.~Nunes,
  Phys.\ Rev.\ D {\bf 94}, no. 12, 123511 (2016)
  [arXiv:1608.02454 [astro-ph.CO]].
  
  \bibitem{Kumar:2017dnp} 
  S.~Kumar and R.~C.~Nunes,
  Phys.\ Rev.\ D {\bf 96}, no. 10, 103511 (2017)
  [arXiv:1702.02143 [astro-ph.CO]].
  
  
  
  \bibitem{DiValentino:2017iww} 
  E.~Di Valentino, A.~Melchiorri and O.~Mena,
  Phys.\ Rev.\ D {\bf 96}, no. 4, 043503 (2017)
  [arXiv:1704.08342 [astro-ph.CO]].
  
   \bibitem{Yang:2018euj} 
  W.~Yang, S.~Pan, E.~Di Valentino, R.~C.~Nunes, S.~Vagnozzi and D.~F.~Mota,
  JCAP {\bf 1809}, no. 09, 019 (2018)
  [arXiv:1805.08252 [astro-ph.CO]].
  
 
  
  
  \bibitem{Yang:2018uae} 
  W.~Yang, A.~Mukherjee, E.~Di Valentino and S.~Pan,
  Phys.\ Rev.\ D {\bf 98}, no. 12, 123527 (2018)
  [arXiv:1809.06883 [astro-ph.CO]].
  
  
  \bibitem{Kumar:2019wfs} 
  S.~Kumar, R.~C.~Nunes and S.~K.~Yadav,
  Eur.\ Phys.\ J.\ C {\bf 79}, no. 7, 576 (2019)
  [arXiv:1903.04865 [astro-ph.CO]].  


   \bibitem{Pan:2019gop} 
  S.~Pan, W.~Yang, E.~Di Valentino, E.~N.~Saridakis and S.~Chakraborty,
Phys. Rev. D \textbf{100}, no.10, 103520 (2019)
  arXiv:1907.07540 [astro-ph.CO]. 
  
    \bibitem{DiValentino:2019ffd} 
  E.~Di Valentino, A.~Melchiorri, O.~Mena and S.~Vagnozzi,
  Phys. Dark Univ. \textbf{30}, 100666 (2020)
  arXiv:1908.04281 [astro-ph.CO].


  \bibitem{DiValentino:2019jae}
E.~Di Valentino, A.~Melchiorri, O.~Mena and S.~Vagnozzi,
Phys. Rev. D \textbf{101}, no.6, 063502 (2020)
[arXiv:1910.09853 [astro-ph.CO]].

\bibitem{Zhai:2023yny}
Y.~Zhai, W.~Giar\`e, C.~van de Bruck, E.~Di Valentino, O.~Mena and R.~C.~Nunes,
JCAP \textbf{07}, 032 (2023)
[arXiv:2303.08201 [astro-ph.CO]].
  
  
  \bibitem{Pourtsidou:2016ico} 
  A.~Pourtsidou and T.~Tram,
  Phys.\ Rev.\ D {\bf 94}, no. 4, 043518 (2016)
  [arXiv:1604.04222 [astro-ph.CO]].
  
 \bibitem{An:2017crg} 
  R.~An, C.~Feng and B.~Wang,
  JCAP {\bf 1802}, no. 02, 038 (2018)
  [arXiv:1711.06799 [astro-ph.CO]].

  \bibitem{Gleyzes:2015pma} 
  J.~Gleyzes, D.~Langlois, M.~Mancarella and F.~Vernizzi,
  JCAP {\bf 1508}, 054 (2015)
  [arXiv:1504.05481 [astro-ph.CO]].
  
  \bibitem{Boehmer:2015kta} 
  C.~G.~B\"{o}ehmer, N.~Tamanini and M.~Wright,
  Phys.\ Rev.\ D {\bf 91}, no. 12, 123002 (2015)
  [arXiv:1501.06540 [gr-qc]].
  
  \bibitem{Boehmer:2015sha} 
  C.~G.~B\"{o}ehmer, N.~Tamanini and M.~Wright,
  Phys.\ Rev.\ D {\bf 91}, no. 12, 123003 (2015)
  [arXiv:1502.04030 [gr-qc]].

\bibitem{vandeBruck:2015ida} 
  C.~van de Bruck and J.~Morrice,
  JCAP {\bf 1504}, 036 (2015)
  [arXiv:1501.03073 [gr-qc]].

\bibitem{Xiao:2018jyl} 
  L.~Xiao, R.~An, L.~Zhang, B.~Yue, Y.~Xu and B.~Wang,
  Phys.\ Rev.\ D {\bf 99}, no. 2, 023528 (2019)
  [arXiv:1807.05541 [astro-ph.CO]].


\bibitem{Amico:2016qft} G. D'Amico, T. Hamill and Nemanja Kaloper,
Phys. Rev. D  {\bf 94}, 103526 (2016)
[arXiv:1605.00996 [hep-th]]. 

\bibitem{Marsh:2017prd} M. C. D. Marsh,
Phys. Rev. Lett.  {\bf 118}, 011302 (2017)
[arXiv:1606.01538 [astro-ph.CO]].

\bibitem{Kase:2019hor} R. Kase and S. Tsujikawa, 
Phys. Rev. D \textbf{101}, no.6, 063511 (2020)
[arXiv:1910.02699 [gr-qc]]. 

 \bibitem{Alexander:2019wne} 
  S.~Alexander, M.~Cort\^{e}s, A.~R.~Liddle, J.~Magueijo, R.~Sims and L.~Smolin,
  Phys. Rev. D \textbf{100}, no.8, 083507 (2019)
  arXiv:1905.10382 [gr-qc].
  
 \bibitem{Pan:2020zza}
S.~Pan, G.~S.~Sharov and W.~Yang,
Phys. Rev. D \textbf{101}, no.10, 103533 (2020)
[arXiv:2001.03120 [astro-ph.CO]]. 

\bibitem{Valiviita:2008iv}
J.~Valiviita, E.~Majerotto and R.~Maartens,
JCAP \textbf{07}, 020 (2008)
[arXiv:0804.0232 [astro-ph]].

\bibitem{Pan:2020mst}
S.~Pan, J.~de Haro, W.~Yang and J.~Amor\'os,
Phys. Rev. D \textbf{101}, no.12, 123506 (2020)
[arXiv:2001.09885 [gr-qc]].

\bibitem{Yang:2021oxc}
W.~Yang, S.~Pan, L.~Arest\'e Sal\'o and J.~de Haro,
Phys. Rev. D \textbf{103}, no.8, 083520 (2021)
[arXiv:2104.04505 [astro-ph.CO]].


\bibitem{Muller:2004yb}
C.~M.~Muller,
Phys. Rev. D \textbf{71}, 047302 (2005)
[arXiv:astro-ph/0410621 [astro-ph]].

\bibitem{Kumar:2012gr}
S.~Kumar and L.~Xu,
Phys. Lett. B \textbf{737}, 244-247 (2014)
[arXiv:1207.5582 [gr-qc]].

\bibitem{Armendariz-Picon:2013jej}
C.~Armendariz-Picon and J.~T.~Neelakanta,
JCAP \textbf{03}, 049 (2014)
[arXiv:1309.6971 [astro-ph.CO]].

\bibitem{Kopp:2018zxp}
M.~Kopp, C.~Skordis, D.~B.~Thomas and S.~Ili\'c,
Phys. Rev. Lett. \textbf{120}, no.22, 221102 (2018)
[arXiv:1802.09541 [astro-ph.CO]].

\bibitem{Ilic:2020onu}
S.~Ili\'c, M.~Kopp, C.~Skordis and D.~B.~Thomas,
Phys. Rev. D \textbf{104}, no.4, 043520 (2021)
[arXiv:2004.09572 [astro-ph.CO]].

\bibitem{Naidoo:2022rda}
K.~Naidoo, M.~Jaber, W.~A.~Hellwing and M.~Bilicki,
[arXiv:2209.08102 [astro-ph.CO]].

\bibitem{Pan:2022qrr}
S.~Pan, W.~Yang, E.~Di Valentino, D.~F.~Mota and J.~Silk,
[arXiv:2211.11047 [astro-ph.CO]].


  \bibitem{DiValentino:2016hlg}
E.~Di Valentino, A.~Melchiorri and J.~Silk,
Phys. Lett. B \textbf{761}, 242-246 (2016)
[arXiv:1606.00634 [astro-ph.CO]].




\bibitem{Mukhanov:1990me}
V.~F.~Mukhanov, H.~A.~Feldman and R.~H.~Brandenberger,
Phys. Rept. \textbf{215}, 203-333 (1992)

\bibitem{Ma:1995ey}
C.~P.~Ma and E.~Bertschinger,
Astrophys. J. \textbf{455}, 7-25 (1995)
[arXiv:astro-ph/9506072 [astro-ph]].

\bibitem{Malik:2008im}
K.~A.~Malik and D.~Wands,
Phys. Rept. \textbf{475}, 1-51 (2009)
[arXiv:0809.4944 [astro-ph]].


\bibitem{ACT:2020frw}
S.~K.~Choi \textit{et al.} [ACT],
JCAP \textbf{12}, 045 (2020)
[arXiv:2007.07289 [astro-ph.CO]].

\bibitem{ACT:2020gnv}
S.~Aiola \textit{et al.} [ACT],
JCAP \textbf{12}, 047 (2020)
[arXiv:2007.07288 [astro-ph.CO]].



\bibitem{Wei:2010fz}
H.~Wei,
Nucl. Phys. B \textbf{845}, 381-392 (2011)
doi:10.1016/j.nuclphysb.2010.12.010
[arXiv:1008.4968 [gr-qc]].

\bibitem{Wei:2010cs}
H.~Wei,
Commun. Theor. Phys. \textbf{56}, 972-980 (2011)
[arXiv:1010.1074 [gr-qc]].

\bibitem{Li:2011ga}
Y.~H.~Li and X.~Zhang,
Eur. Phys. J. C \textbf{71}, 1700 (2011)
[arXiv:1103.3185 [astro-ph.CO]].

\bibitem{Guo:2017deu}
J.~J.~Guo, J.~F.~Zhang, Y.~H.~Li, D.~Z.~He and X.~Zhang,
Sci. China Phys. Mech. Astron. \textbf{61}, no.3, 030011 (2018)
[arXiv:1710.03068 [astro-ph.CO]].

\bibitem{Arevalo:2019axj}
F.~Arevalo, A.~Cid, L.~P.~Chimento and P.~Mella,
Eur. Phys. J. C \textbf{79}, no.4, 355 (2019)
[arXiv:1901.04300 [gr-qc]].






\bibitem{Teixeira:2023zjt}
E.~M.~Teixeira, R.~Daniel, N.~Frusciante and C.~van de Bruck,
[arXiv:2309.06544 [astro-ph.CO]].

\bibitem{Gomez-Valent:2022bku}
A.~G\'omez-Valent, Z.~Zheng, L.~Amendola, C.~Wetterich and V.~Pettorino,
Phys. Rev. D \textbf{106}, no.10, 103522 (2022)
[arXiv:2207.14487 [astro-ph.CO]].

\bibitem{Gomez-Valent:2020mqn}
A.~G\'omez-Valent, V.~Pettorino and L.~Amendola,
Phys. Rev. D \textbf{101}, no.12, 123513 (2020)
[arXiv:2004.00610 [astro-ph.CO]].

\bibitem{VanDeBruck:2017mua}
C.~Van De Bruck and J.~Mifsud,
Phys. Rev. D \textbf{97}, no.2, 023506 (2018)
[arXiv:1709.04882 [astro-ph.CO]].

\bibitem{vandeBruck:2016hpz}
C.~van de Bruck, J.~Mifsud and J.~Morrice,
Phys. Rev. D \textbf{95}, no.4, 043513 (2017)
[arXiv:1609.09855 [astro-ph.CO]].



\bibitem{Wang:2018azy}
Y.~Wang and G.~B.~Zhao,
Astrophys. J. \textbf{869}, no.1, 26 (2018)
[arXiv:1805.11210 [astro-ph.CO]].

\bibitem{Yang:2019uzo} 
  W.~Yang, O.~Mena, S.~Pan and E.~Di Valentino,
  Phys. Rev. D \textbf{100},  083509 (2019)
  arXiv:1906.11697 [astro-ph.CO].


\bibitem{Yang:2020tax}
W.~Yang, E.~Di Valentino, O.~Mena and S.~Pan,
Phys. Rev. D \textbf{102}, no.2, 023535 (2020)
[arXiv:2003.12552 [astro-ph.CO]].

\bibitem{Yao:2022kub}
Y.~H.~Yao and X.~H.~Meng,
[arXiv:2207.05955 [astro-ph.CO]].

\bibitem{Yang:2022csz}
W.~Yang, S.~Pan, O.~Mena and E.~Di Valentino,
[arXiv:2209.14816 [astro-ph.CO]].




\bibitem{Gaztanaga:2008de}
E.~Gaztanaga, R.~Miquel and E.~Sanchez,
Phys. Rev. Lett. \textbf{103}, 091302 (2009)
[arXiv:0808.1921 [astro-ph]].

\bibitem{Mortonson:2009nw}
M.~J.~Mortonson,
Phys. Rev. D \textbf{80}, 123504 (2009)
[arXiv:0908.0346 [astro-ph.CO]].

\bibitem{Suyu:2013kha}
S.~H.~Suyu, T.~Treu, S.~Hilbert, A.~Sonnenfeld, M.~W.~Auger, R.~D.~Blandford, T.~Collett, F.~Courbin, C.~D.~Fassnacht and L.~V.~E.~Koopmans, \textit{et al.}
Astrophys. J. Lett. \textbf{788}, L35 (2014)
[arXiv:1306.4732 [astro-ph.CO]].

\bibitem{LHuillier:2016mtc}
B.~L'Huillier and A.~Shafieloo,
JCAP \textbf{01}, 015 (2017)
[arXiv:1606.06832 [astro-ph.CO]].

\bibitem{DiValentino:2019qzk}
E.~Di Valentino, A.~Melchiorri and J.~Silk,
Nature Astron. \textbf{4}, no.2, 196-203 (2019)
[arXiv:1911.02087 [astro-ph.CO]].

\bibitem{Handley:2019tkm}
W.~Handley,
Phys. Rev. D \textbf{103}, no.4, L041301 (2021)
[arXiv:1908.09139 [astro-ph.CO]].



\bibitem{DiValentino:2020kpf}
E.~Di Valentino, A.~Melchiorri, O.~Mena, S.~Pan and W.~Yang,
Mon. Not. Roy. Astron. Soc. \textbf{502}, no.1, L23-L28 (2021)
[arXiv:2011.00283 [astro-ph.CO]].

\bibitem{Yang:2021hxg}
W.~Yang, S.~Pan, E.~Di Valentino, O.~Mena and A.~Melchiorri,
JCAP \textbf{10}, 008 (2021)
[arXiv:2101.03129 [astro-ph.CO]].


\end{thebibliography}
\end{document}